\documentclass[12pt,preprint]{aastex}

\slugcomment{Astrophysical Journal}

\shorttitle{Nuclear Starbursts in Seyfert 2s}
\shortauthors{Imanishi}

\begin{document}

\title{Compact Nuclear Starbursts in Seyfert 2 Galaxies from the CfA and
12$\mu$m Samples}

\author{Masatoshi Imanishi\altaffilmark{1,2,3}}
\affil{National Astronomical Observatory, Mitaka, 2-21-1, Osawa, Tokyo
181-8588, Japan} 
\email{imanishi@optik.mtk.nao.ac.jp} 

\altaffiltext{1}{Visiting Astronomer at the Infrared Telescope Facility,
which is operated by the University of Hawaii under
Cooperative Agreement no. NCC 5-538 with the National
Aeronautics and Space Administration, Office of Space
Science, Planetary Astronomy Program.}

\altaffiltext{2}{Based in part on data collected at Subaru Telescope,
which is operated by the National Astronomical Observatory of Japan.}

\altaffiltext{3}{Department of Astronomy, School of Science, Graduate 
University for Advanced Studies, Mitaka, Tokyo 181-8588}

\begin{abstract}
We present infrared 2.8--4.1 $\mu$m slit spectra of 32 Seyfert 2
galaxies in the CfA and 12 $\mu$m samples.   
The 3.3 $\mu$m polycyclic aromatic hydrocarbon (PAH) emission feature 
was used to estimate the absolute magnitude of a compact nuclear
starburst (less than a few hundred parsecs in size) that is presumed to
have occurred in the outer region of an obscuring dusty molecular torus
around a central supermassive black hole.
We detected 3.3 $\mu$m PAH emission in 11 of the 32 Seyfert 2 nuclei in
our sample, providing evidence for the presence of compact nuclear
starbursts in a significant fraction of Seyfert 2 nuclei.  
However, the rest-frame equivalent widths of the 3.3 $\mu$m PAH
emission, and the 3.3 $\mu$m PAH-to-infrared luminosity ratios measured
in this study suggest that compact nuclear starbursts generally do not
contribute significantly to the observed 3--4 $\mu$m nuclear fluxes or
to the infrared luminosities of Seyfert 2 galaxies. 
Absorption features at 3.4 $\mu$m from bare dust were clearly detected
in only two of the nuclei, and features at 3.1 $\mu$m from ice-covered
dust were detected in only one nucleus. 
If the dust properties in the direction of these Seyfert 2 nuclei do not
differ significantly from the Galactic interstellar medium, then these
small absorption optical depths suggest that dust extinction toward the
3--4 $\mu$m continuum-emitting region in the innermost part of the
obscuring dusty torus is modest: A$_{\rm V}$ $<$ 50--60 mag.   
Finally, the 3.3 $\mu$m PAH emission luminosities measured in this study
were found to be significantly correlated with {\it IRAS} 12- and
25-$\mu$m, and nuclear $N$-band (10.6 $\mu$m) luminosities. 
If these three luminosities trace the power of the active galactic
nucleus (AGN), then the luminosities of compact
nuclear starbursts and AGNs are correlated.   
This correlation is in agreement with theories predicting that the
presence of a compact nuclear starburst in the torus leads to an
enhancement of the mass accretion rate onto the central supermassive
black hole.

\end{abstract}


\section{Introduction}

Seyferts are the most numerous class of active galactic nuclei (AGNs). 
The two types of Seyfert galaxies, type 1 (with broad optical emission
lines) and type 2 (with no such lines), can be unified by the
combination of an obscuring dusty molecular torus around a central,
accreting supermassive black hole, and different viewing angles 
\citep{ant93}.  
As dusty tori are rich in molecular gas, they are natural sites for
starbursts to occur \citep{fab98}. 
For several reasons, such a {\it nuclear} starburst in the torus is
likely to occur at the outer part, as originally proposed by
\citet{hec97}.  
First, for a dusty molecular torus whose mass is significantly smaller
than the central supermassive black hole \citep{tan98}, 
gas in the torus rotates in a Keplerian manner, and the gravitational
stability parameter Q \citep{too64} is described as 
Q = $\sigma$$\Omega$/$\pi$G$\Sigma_{\rm gas}$, 
where $\sigma$, $\Omega$, and $\Sigma_{\rm gas}$ are the gas velocity
dispersion, angular velocity, and gas surface density, respectively. 
In the case of a power-law radial density distribution of gas
($\Sigma_{\rm gas}$ $\propto$ r$^{-\alpha}$; where $r$ is the distance
from the central supermassive black hole) in an isothermal Keplerian
disk ($\Omega$ $\propto$ r$^{-1.5}$), Q is proportional to r$^{\alpha-1.5}$. 
If the radial density distribution of gas and dust in the torus is
flatter than r$^{-1.5}$, as is often assumed
\citep{bar87,gra94,row95,efs95}, then Q decreases with radius, so
that the gravitational collapse of molecular gas can occur more easily
in the outer regions of the torus \citep{wad02}.
In addition, star formation in the inner part of the torus could be
suppressed due to strong X-ray radiation from the AGN \citep{cid95}.  
Finally, if there is molecular gas inflow from the host galaxy to
the torus, gas compression will occur at the outer edges of the torus, 
encouraging star formation. 

Such compact nuclear starbursts have been detected in some Seyfert nuclei
\citep{gon98,oli99,sto00,gon01,cid01,koh02}.   
However, their absolute luminosities and relation to the central AGNs are
still unclear. 
Infrared
2.8--4.1 $\mu$m spectroscopy is one of the most
powerful ways of detecting and quantitatively estimating the energetic
importance of these compact nuclear starbursts \citep{ima02,rod03}.    
First, starburst emission is clearly distinguishable from emission from
the AGN, using the polycyclic aromatic hydrocarbon (PAH) emission found
at 3.3 $\mu$m.
In a {\it normal} starburst, PAHs widely distributed in the
interstellar medium are excited by far-UV photons from stars, and 
so strong 3.3 $\mu$m PAH emission is usually detected 
\citep{moo86}. 
Close to an AGN, where sufficient far-UV photons from
the AGN are available, X-ray radiation is also strong, which destroys 
PAHs \citep{voi92}.  
At a certain distance from the central AGN in the obscuring material, 
the X-ray flux is sufficiently attenuated to prevent this. 
However, as UV photons have a higher susceptibility to extinction than
X-rays, far-UV photons from the AGN are not expected to survive to these
distances. 
Thus, no PAH emission is expected in a {\it pure} AGN. 
The detection of 3.3 $\mu$m PAH emission from an AGN requires far-UV
photons from a starburst to be sufficiently shielded from AGN emission,
as is the case for a compact nuclear starburst occurring in the outer
part of a dusty torus. 
The PAH emission luminosity is therefore expected to roughly trace the
magnitude of such a starburst. 
Another advantage of infrared 2.8--4.1 $\mu$m spectroscopy is that the
flux attenuation is lower, as dust extinction at 2.8--4.1 $\mu$m is much
less than at UV or optical wavelengths.
The absolute magnitude of a starburst in a
Seyfert can be determined from the {\it observed} 3.3 $\mu$m
PAH emission luminosity \citep{ima02}.  
Finally, since 3.3 $\mu$m PAH emission is intrinsically strong
\citep{imd00}, even the signature of a weak starburst is detectable with a 
spectrum of typical signal-to-noise ratio.  
Using infrared 2.8--4.1 $\mu$m slit spectroscopy, it has been possible to 
detect and quantify previously undetected compact nuclear
starbursts in Seyferts \citep{ima02,rod03}.
           
The physical scale of a compact nuclear starburst in the outer
region of a dusty molecular torus is less than a few hundred parsecs,
which corresponds to less than 1--2 arcsec in the majority of Seyferts
at $z >$ 0.007.  
In Seyferts, extended {\it circumnuclear} ring-shaped starbursts are
found typically at distances of $\sim$1 kpc from the nuclei
\citep{sto96a,sto96b}, and are generally more powerful than the compact
{\it nuclear} starbursts \citep{lef01,ima02}. 
Hence, the use of a narrow ($<$1--2 arcsec) slit is most effective to
distinguish between extended starbursts and the compact nuclear
starbursts that are of interest. 
Therefore, we have performed ground-based infrared 2.8--4.1 $\mu$m slit
spectroscopy focusing initially on Seyfert 2 nuclei to determine
observational constraints on the properties of compact nuclear starbursts.  
Throughout this paper, $H_{0}$ $=$ 75 km s$^{-1}$ Mpc$^{-1}$,
$\Omega_{\rm M}$ = 0.3, and $\Omega_{\rm \Lambda}$ = 0.7 are adopted.

\section{Targets}

Seyfert 2 galaxies in the CfA \citep{huc92} and 12$\mu$m (Rush, Malkan,
\& Spinoglio 1993) samples were selected based on their host galaxy magnitudes
and {\it IRAS} 12 $\mu$m fluxes, respectively. 
These samples are not expected to be biased toward or against the
presence of compact nuclear starbursts, and is thus suitable for
our program.  
The following sample selection criteria were employed: 
(1) In very nearby sources, slit spectroscopy may miss a significant
fraction of emission from compact nuclear starbursts with a physical
size scale of up to a few hundred pc. 
The slit width used, 0$\farcs$9--1$\farcs$6 (\S 3), corresponds to
120--220 pc at $z=0.007$.  
Seyfert 2s at $z$ $<$ 0.007 are thus excluded.
(2) In distant Seyfert 2s, slit spectroscopy may be significantly
contaminated by extended (kpc scale) powerful circumnuclear starbursts. 
The slit width used corresponds to 600 pc -- 1 kpc at $z=0.035$.
Seyfert 2s at $z$ $>$ 0.035 are excluded. 
(3) 12 $\mu$m Seyfert 2s at declination $<-$35$^{\circ}$ are excluded
because they are not observable under good observing conditions
from Mauna Kea, Hawaii.  
(4) Some 12$\mu$m Seyfert 2s in \citet{rus93} have been later re-classified
as LINERs or HII-region galaxies based on higher quality optical
spectra \citep{tra03}. These sources are excluded.  
All CfA Seyfert 2s later re-observed by \citet{ost93} were 
re-classified as Seyfert 1.8, 1.9, or 2. 
These sources are included. 

These selection criteria result in 18 CfA Seyfert 2s and 26 12$\mu$m
Seyfert 2s, of which six sources are included in both samples. 
From the total sample of 38 Seyfert 2s, 32 (32/38 = 84\%) sources have
been observed so far.    
Table~\ref{tbl-1} summarizes the infrared emission properties of the 32
observed Seyfert 2s.   
The six unobserved sources are  UGC 1395 (=0152+06), NGC 3362, UGC 6100 
(=A1058+45), UGC 8621 (=1335+39), and NGC 5283 in the CfA sample, and
MCG-2-8-39 in the 12$\mu$m sample.  
Although the sample is not statistically complete, major conclusions
drawn from the 32 sources in this paper are unlikely to be affected
significantly by omission of the six unobserved sources. 

\section{Observations and Data Analysis}
 
The observing log is shown in Table~\ref{tbl-2}.
Observations were made with the UKIRT CGS4 \citep{moun90}, the IRTF NSFCAM
\citep{shu94}, the IRTF SpeX \citep{ray03}, and the Subaru IRCS \citep{kob00},
on Mauna Kea, Hawaii.
Details of the observations with the UKIRT CGS4 (for NGC 7172) and the
IRTF NSFCAM (for NGC 5256 and NGC 5135) were described by \citet{ima00} and
\citet{ima02}, respectively, and are not repeated here.

For the IRTF SpeX observing run, the 1.9--4.2 $\mu$m cross-dispersed mode
with a 1\farcs6 wide slit was employed. 
The achievable spectral resolution at 3.5 $\mu$m is R $\sim$ 450.   
The sky conditions were photometric and the seeing sizes at $K$ (2.2 
$\mu$m) were 0$\farcs$6--0$\farcs$9 (full-width at half-maximum; FWHM)
throughout the observations.  
The position angle of the slit was set along the north-south direction. 
A standard telescope nodding technique (ABBA pattern) with a throw of 7.5
arcsec was employed along the slit to subtract background emission.
As 3--4 $\mu$m emission from a Seyfert 2 galaxy is usually dominated
by compact nuclear emission \citep{alo98}, this throw is believed to be
sufficiently large.
The telescope tracking was monitored with the infrared slit-viewer of SpeX.  
Each exposure was 15 sec, and 2 coadds were
made at each position. 
With this exposure time, signals at $\lambda_{\rm obs}$ $>$ 4.1 $\mu$m
(in the observed frame) exceed the linearity level of the SpeX array, and so
data at $\lambda_{\rm obs}$ $>$ 4.1 $\mu$m were removed.  

For the Subaru IRCS observing run, a 0$\farcs$9-wide slit and the
$L$-grism were used with a 58-mas pixel scale. 
The achievable spectral resolution at 3.5 $\mu$m is R $\sim$ 140.  
The sky was not completely photometric during the observations of some
of the Seyfert 2 nuclei (Mrk 334, NGC 1144, and NGC 1125).  
The position angle of the slit was set along the east-west direction, except
in the case of NGC 1144 where the angle was set along the north-south
direction.  
The seeing at $K$ was 0$\farcs$5--0$\farcs$8 in FWHM. 
A standard telescope nodding technique (ABBA pattern), with a throw of 7
arcsec along the slit, was employed. 
Each exposure was 2.0--3.0 sec, and 15--30 coadds were made at each slit
position.  
Data at $\lambda_{\rm obs}$ $>$ 4.1 $\mu$m were removed, due to signal
levels above the linearity level of the IRCS array.   

A-, F-, and G-type standard stars (Table~\ref{tbl-2}) 
were observed with airmass difference of $<$0.1 to the individual
Seyfert 2 nuclei, to correct for the transmission of the Earth's
atmosphere. 
The $L$-band (3.5 $\mu$m) magnitudes of standard stars were estimated
from their $V$-band (0.6 $\mu$m) magnitudes, adopting the $V-L$
colors appropriate to the stellar types of individual standard stars
\citep{tok00}.

Standard data analysis procedures were employed using IRAF
%--------------
\footnote{
IRAF is distributed by the National Optical Astronomy Observatories,
which are operated by the Association of Universities for Research
in Astronomy, Inc. (AURA), under cooperative agreement with the
National Science Foundation.}.
%--------------
Initially, bad pixels and pixels hit by cosmic rays were replaced with the
interpolated values of the surrounding pixels.
Then, frames taken with an A (or B) beam were subtracted from
frames subsequently taken with a B (or A) beam, and
the resulting subtracted frames were added and were divided by a
spectroscopic flat image. 
The Seyfert 2 and standard star spectra were then
extracted by integrating signals over 2$\farcs$4--3$\farcs$0 and
1$\farcs$4--1$\farcs$8 along the slit for the IRTF SpeX and Subaru IRCS
data, respectively.  
Wavelength calibration was performed using the wavelength-dependent
transmission of the Earth's atmosphere. 
The Seyfert 2 spectra were divided by those of the corresponding standard
stars, and were multiplied by the spectra of blackbodies with
temperatures corresponding to those of the individual standard stars
(Table~\ref{tbl-2}). 

All of the IRTF SpeX observing runs and
Subaru IRCS runs in 2003 March were made under photometric sky
conditions, and so flux calibration was made based on the signal
detected inside our slit. 
For the CfA sample Seyfert 2s, $L$-band (3.5 $\mu$m) photometry taken with
1$\farcs$5--8$\farcs$6 apertures was available \citep{iva00,alo03}.     
Our flux calibration generally agrees with or is slightly ($<$1 mag) fainter
than their photometry, which is reasonable because 3--4 $\mu$m continuum
emission from a Seyfert 2 galaxy is usually dominated by compact nuclear
emission, but emission more extended than our slit size also exists
\citep{alo98,alo01,alo03}.     
During the observations of Mrk 334, NGC 1125, and NGC 1144, the sky was
not completely photometric due to the presence of thin cirrus.  
For Mrk 334 and NGC 1144, $L$-band photometry made with a 1.5 arcsec
aperture by \citet{alo03} was used for the flux calibration of our slit
spectra.  
For NGC 1125, although no $L$-band photometry was available, 
the cirrus was thinner than during the observations of Mrk 334 and
NGC 1144, and the flux discrepancy among independent data sets 
is small ($<$20\%). 
The flux was thus calibrated using our slit spectra. 

Appropriate spectral binning was applied for faint sources, particularly
at $\lambda_{\rm obs}$ $<$ 3.3 $\mu$m. 
In this wavelength range, the Earth's atmospheric transmission is highly
wavelength dependent, and even displacement at the sub-pixel level between a
Seyfert 2 and corresponding standard star along the wavelength direction
could produce a spiky spectrum. 
Binning to a spectral resolution of R $<$ 100 can help reduce these
spurious spikes \citep{ima03}. 
The targeted feature in this wavelength range is the 3.1 $\mu$m
absorption feature caused by ice-covered dust grains, which is
sufficiently broad (Smith, Sellgren, \& Tokunaga 1989) that a spectral
resolution of R $>$ 50 allows investigation of its properties.     
In the vicinity of the redshifted 3.3 $\mu$m PAH emission feature
($\lambda_{\rm rest}$ $\sim$ 3.29 $\mu$m), a spectral resolution of R
$>$ 100 was retained to properly trace its profile and estimate its flux
reliably.   
For faint sources, data at the longer side of 3.3 $\mu$m PAH emission
were also binned. 
The main targeted feature in this wavelength range is the absorption
feature due to bare carbonaceous dust at $\lambda_{\rm rest}$ $\sim$
3.4 $\mu$m \citep{imd00,idm01}.  
As this feature is broad and weak, a spectral resolution of R $>$ 100 is
sufficient.  

\section{Results}

Figure~\ref{fig1} shows flux-calibrated 2.8--4.1 $\mu$m slit spectra. 
For NGC 5256 and NGC 5135, the 3.0--4.0 $\mu$m spectra were presented 
by \citet{ima02}. 
Here, 2.8--4.1 $\mu$m spectra are shown, which illustrate the properties
of the broad 3.1 $\mu$m absorption feature. The spectrum of NGC 7172 was
taken with UKIRT CGS4, so the wavelength coverage is $\lambda_{\rm obs}$
= 3.2--3.8 $\mu$m.  
The spectrum presented by \citet{ima00} is shown again here. 

In the 2.8--4.1 $\mu$m spectra shown in Figure~\ref{fig1}, we consider
3.3 $\mu$m PAH emission to be detected when at least two successive data
points are significantly higher than the continuum level. 
Using this criterion, 3.3 $\mu$m PAH emission was detected in 11 of the
32 observed Seyfert 2 nuclei.  
The detection rates are 31\% (4/13) and 36\% (9/25) for CfA and 12 $\mu$m
Seyfert 2s, respectively.  
The fluxes, luminosities, and rest-frame equivalent widths of the
3.3 $\mu$m PAH emission were estimated using the method described by
\citet{ima02} and are summarized in Table~\ref{tbl-3}. 

If sufficiently obscured AGN emission contributes significantly to an
observed 3--4 $\mu$m nuclear flux, the 3.4 $\mu$m absorption feature due
to bare carbonaceous dust grains \citep{imd00,idm01} and/or the 3.1
$\mu$m absorption feature due to ice-covered dust grains \citep{ima03}
should be detected.   
The 3.4 $\mu$m absorption feature was clearly detected in only 
F04385$-$0828 and NGC 7172, with an optical depth of $\sim$0.1.   
The presence of the 3.1 $\mu$m absorption feature was inferred only in NGC
4388.   

In some bright nearby sources (NGC 4388, NGC 262, NGC 1194, NGC 4968
MCG-3-34-64, and MCG-2-40-4) for which sufficient signal-to-noise
ratios in the continuum are achieved even without spectral binning,
[SiIV] emission at $\lambda_{\rm rest}$ = 3.94 $\mu$m and Br$\alpha$
emission at $\lambda_{\rm rest}$ = 4.05 $\mu$m are detected. 
These emission lines are sufficiently strong in NGC 4388 and NGC 4968
that reliable flux estimates for these relatively narrow emission
lines (narrower than PAH) are possible even in our low-resolution spectra. 
For NGC 4388, the observed [SiIX] and Br$\alpha$ emission fluxes are
estimated to be 3 $\times$ 10$^{-17}$ W m$^{-2}$ and 4 $\times$ 10$^{-17}$  
W m$^{-2}$, respectively.  
For NGC 4968, they are 1 $\times$ 10$^{-17}$ W m$^{-2}$ and 2 $\times$
10$^{-17}$ W m$^{-2}$, respectively.
These fluxes are slightly higher than the estimates by \citet{lut02},
possibly due to our larger slit size.

\section{Discussion}

\subsection{Compact Nuclear Starbursts}

To investigate the contribution from compact nuclear
starbursts to the observed 3--4 $\mu$m fluxes and to the total infrared 
(8--1000 $\mu$m) luminosities, the rest-frame equivalent widths of 3.3
$\mu$m PAH emission features (EW$_{\rm 3.3PAH}$) and 3.3 $\mu$m
PAH-to-infrared luminosity ratios (L$_{\rm 3.3PAH}$/L$_{\rm IR}$) are
shown in Table~\ref{tbl-3}, as described in \citet{ima02}.
 
As an equivalent width ($\equiv$ the ratio of line to continuum flux)
is, by definition, robust to dust extinction, 
a large PAH equivalent width ($\sim$100 nm; Imanishi \& Dudley 2000)
should always be detected in a starburst galaxy, regardless of dust
extinction.   
All but NGC 5256 and Mrk 938 show EW$_{\rm 3.3PAH}$
more than a factor of two smaller than $\sim$100 nm. 
Therefore, although the compact nuclear starbursts are detected
at 3--4 $\mu$m in $\sim$30\% (11/32) of Seyfert 2s, their observed 3--4 $\mu$m
fluxes, except NGC 5256 and Mrk 938, should be dominated by a continuum
that does not show a strong 3.3 $\mu$m PAH emission feature.
AGN-powered hot ($\sim$800--1000K) dust
in the very inner part of an obscuring dusty torus \citep{bar87} close
to the innermost dust sublimation radius (which corresponds to
1000--1800 K dust; Granato et al. 1997) can emit 3--4 $\mu$m continuum
efficiently (assuming a blackbody spectrum), without strong 3.3 $\mu$m
PAH emission.  
Thus, it can be concluded that the 3--4 $\mu$m fluxes in
the majority of the observed Seyfert 2 nuclei originate in featureless
continuum emission from hot dust powered by an AGN. 

The L$_{\rm 3.3PAH}$/L$_{\rm IR}$ ratios in all Seyfert 2s 
(Table~\ref{tbl-3}) are a factor of 
${^{\displaystyle >}_{\displaystyle \sim}}$2 smaller than the ratios for
starburst-dominated galaxies ($\sim$1 $\times$ 10$^{-3}$; Mouri et
al. 1990; Imanishi 2002). 
Thus, compact nuclear starbursts are unlikely to dominate the
infrared luminosities of these Seyfert 2 galaxies.     
The larger sample presented here confirms the small EW$_{\rm 3.3PAH}$
and L$_{\rm 3.3PAH}$/L$_{\rm IR}$  
found previously in a smaller sample of Seyfert 2s \citep{ima02}. 

\subsection{Dust Extinction toward the Innermost 3--4 $\mu$m
Continuum-emitting Regions in Seyfert 2 Nuclei} 

As it has been shown in $\S$5.1 that the observed 3--4 $\mu$m fluxes
in the majority of the observed Seyfert 2 nuclei are dominated
by AGN-powered hot-dust emission in the innermost part of the obscuring
dusty torus, the observed optical depths of the absorption features at
3.4 $\mu$m and 3.1 $\mu$m reflect the column density of obscuring dust
toward the innermost 3--4 $\mu$m continuum-emitting regions.
In the Galaxy, dust grains located deep inside molecular gas are covered
with an ice mantle, if they are sufficiently shielded from ambient UV
radiation \citep{whi88}. 
Such dust grains show an absorption feature at 3.1 $\mu$m \citep{smi89}, 
but no significant feature at 3.4 $\mu$m \citep{men01}.
Dust absorption at 3.4 $\mu$m is detected in the diffuse
interstellar medium \citep{pen94,ima96,raw03} where dust grains are not
covered with an ice mantle (that is, ``bare''). 
As dust extinction toward the innermost 3--4 $\mu$m continuum-emitting
hot dust around an AGN can be fitted with a foreground-screen dust 
model, the total column density of obscuring dust, including both bare
and ice-covered dust grains, can be estimated in a fairly
straightforward manner 
from the optical depths of these absorption features ($\tau_{3.1}$ and
$\tau_{3.4}$).  

The $\tau_{3.4}$ values for the two detected sources (F04385$-$0828 and
NGC 7172) are $\sim$0.1.  
Conservative upper limits for $\tau_{3.4}$ in the remaining
undetected sources are $<$0.2. 
The $\tau_{3.1}$ value for NGC 4388 is $\sim$0.2, some fraction of which
may originate in dust in the edge-on host galaxy \citep{pog02}, rather
than the dusty torus.  
Conservative upper limits for $\tau_{3.1}$ for the remaining sources
are $<$0.3. 
If the obscuring dust toward these Seyfert 2 nuclei shows a dust
extinction curve 
similar to that of the Galaxy ($\tau_{3.4}$/A$_{\rm V}$ $=$ 0.004--0.007;
Pendleton et al. 1994, $\tau_{3.1}$/A$_{\rm V}$ $\sim$ 0.06; Smith,
Sellgren, \& Brooke 1993; Tanaka et al. 1990; Murakawa, Tamura, \& Nagata
2000), then the $\tau_{3.1}$ and $\tau_{3.4}$ values imply that 
A$_{\rm V}$ $<$ 50 mag and $<$ 5 mag for bare and ice-covered dust
grains, respectively.     
Hence, the total dust column density toward the innermost 3--4 $\mu$m
continuum-emitting hot dust around a central AGN is modest, A$_{\rm V}$
$<$ 50--60 mag, in the majority of Seyfert 2 nuclei. 

If the $\tau_{3.4}$/A$_{\rm V}$ ratios toward these Seyfert 2 nuclei are 
substantially smaller than the Galactic ratio, then the A$_{\rm V}$
value estimated in this way is underestimated in Seyfert 2s. 
In the dusty torus around an AGN, the dust size may be larger than in the
Galactic interstellar medium, due to dust coagulation
\citep{mai01a,mai01b,ima01}, and the 9.7 $\mu$m silicate dust feature
may be weakened \citep{lao93}.  
It is possible that the 3.4 $\mu$m absorption feature is also
weakened and the $\tau_{3.4}$/A$_{\rm V}$ ratio toward an obscured AGN
is smaller than the Galactic value.  
However, in a number of obscured AGNs, it has been found that the dereddened
AGN luminosity based on the Galactic $\tau_{3.4}$/A$_{\rm V}$ ratio
is comparable to the bolometric luminosity \citep{idm01,ima03}. 
If $\tau_{3.4}$/A$_{\rm V}$ ratios in these obscured AGNs are 
substantially smaller than the Galactic ratio, then the dereddened AGN 
luminosities would substantially exceed the bolometric luminosities. 
It is thus unlikely that the $\tau_{3.4}$/A$_{\rm V}$ ratio toward 
an obscured AGN is substantially smaller than the Galactic value, so that
our estimate for dust extinction in these Seyfert 2s appears reasonable.   
Indeed, the estimated modest dust extinction toward the majority of
Seyfert 2 nuclei, excluding the dustiest, most infrared-luminous examples
(ultraluminous infrared galaxies), is generally in good agreement with
other independent estimates.  
First, infrared spectral energy distributions of Seyferts show
similarly modest dust extinction \citep{gra97,dop98,fad98}. 
In addition, Alonso-Herrero, Ward, \& Kotilainen (1997) found that dust
extinction toward the innermost 3.5 $\mu$m continuum-emitting regions 
is A$_{\rm V}$ $<$ 50--60 mag, based on a comparison of the
luminosities of optical [OIII] emission lines from unobscured regions
and 3.5 $\mu$m continuum.  
Finally, \citet{cla00} showed that the equivalent widths of 7.7 $\mu$m
PAH emission measured with {\it ISO} using a large aperture (24 
$\times$ 24 arcsec$^{2}$) are $\sim$8 times larger in Seyfert 2s than
in Seyfert 1s.   
If the 7.7 $\mu$m PAH emission luminosities, mostly originating from
extended starbursts in the host galaxies, are intrinsically the same
in Seyfert 1s and 2s, then the 
larger equivalent widths in Seyfert 2s can be explained by the
attenuation of the 7--8 $\mu$m continuum flux from the AGN.  
Adopting the dust extinction curve with A$_{7-8 \mu m}$/A$_{\rm V}$
$\sim$ 0.04--0.05 derived by \citet{lut96}, the dust extinction toward
7--8 $\mu$m continuum-emission regions in Seyfert 2s is estimated to be 
A$_{\rm V}$ = 45--55 mag.
In fact, the extended PAH emission luminosities in Seyfert 2s are likely to 
be intrinsically larger than in Seyfert 1s \citep{mai95}, so that 
the above dust extinction is probably an overestimate. 
Although the dust extinction toward the innermost 3--4 $\mu$m
continuum-emitting region around an AGN should be larger than that
toward the outer 7--8 $\mu$m continuum-emitting region, we find no
significant difference.  

If the Galactic N$_{\rm H}$/A$_{\rm V}$ ratio is assumed 
($\sim$2 $\times$ 10$^{21}$ cm$^{-2}$ mag$^{-1}$; Predehl \& Schmitt 1995), 
dust extinction of A$_{\rm V}$ $<$ 50--60 mag implies a hydrogen column
density of N$_{\rm H}$ $<$ 10$^{23}$ cm$^{-2}$ in the majority of
Seyfert 2 nuclei, which is substantially smaller than measured values of
N$_{\rm H}$ by X-rays toward AGNs in Seyfert 2 nuclei  
\citep{ris99,bas99}. 
Imaging observations have indicated substantially larger ratios of
N$_{\rm H}$ (columns of X-ray absorbing
material)-to-A$_{\rm V}$ (dust columns toward the innermost 3.5 $\mu$m 
continuum emitting regions) in Seyfert 2s \citep{alo97,alo01,alo03}. 
Our 2.8--4.1 $\mu$m spectroscopy provided additional support for this. 

\subsection{Do More Powerful AGNs Have More Powerful Compact Nuclear
Starbursts?}

\citet{tra01,tra03} argued that there are two distinct types of Seyfert 2
nuclei: one AGN-powered and another powered by a compact nuclear 
starburst. However, \citet{gu02} did not find evidence for such distinct
Seyfert 2 populations. 
The L$_{\rm 3.3PAH}$ values measured from our slit spectra 
(Table~\ref{tbl-3}) are a good measure of the absolute magnitude of
the compact nuclear starbursts \citep{ima02}, and {\it IRAS} 12 $\mu$m
and 25 $\mu$m fluxes are often taken to be representative of the AGN
power in a Seyfert 2 galaxy \citep{spi89,gon01,rod97}.
Ground-based $N$-band (10.6 $\mu$m) aperture photometry of Seyfert 2
nuclei using a single-element bolometer can also be a good tracer of
AGN power \citep{alo02} as the contamination from extended
starbursts is reduced. 

In Figure~\ref{fig2}, we compare L$_{\rm 3.3PAH}$ with 
{\it IRAS} 12- and 25-$\mu$m, and nuclear $N$-band fluxes 
(Table~\ref{tbl-1}). 
Using the generalized Kendall's rank correlation statistic \citep{iso86}
%--------------
\footnote{
software is available at: 
http://www.astro.psu.edu/statcodes/.} 
%--------------
we found the probability that a correlation is not present to be 0.007,
0.002, and 0.001 for Fig.~\ref{fig2}a, ~\ref{fig2}b, and ~\ref{fig2}c,
respectively.  
Thus, we suggest that the luminosities of AGNs and compact nuclear
starbursts are correlated in Seyfert 2 galaxies, so that more powerful AGNs
tend to possess more powerful compact nuclear starbursts in the torus.  

It has been predicted that such compact nuclear starbursts in the dusty
torus can increase the mass accretion rate onto the central supermassive
black hole through enhancement of the turbulence of molecular gas
in the torus \citep{von93,wad02} and/or through radiation effects
\citep{ume97,ume98,ohs99}.    
The luminosity correlation between AGNs and compact nuclear
starbursts is at least in qualitative agreement with these
theoretical predictions.  

However, extended star formation, including circumnuclear ring-shaped
starbursts and quiescent star formation in the host galaxies, may
contribute to the {\it IRAS} 12 $\mu$m and 25 $\mu$m fluxes as well as
to measurements from previous ground-based $N$-band photometry taken
with a single element bolometer with $>$ 3 arcsec apertures. 
$N$-band photometry of spatially-unresolved emission using recently
available two-dimensional large-format mid-infrared arrays at large
telescopes \citep{kra01} is undoubtedly a more reliable AGN indicator.  
These photometric data, if available for our sample in the future, could
be used to investigate the connection between AGNs and compact nuclear
starbursts in much greater detail.  

\subsection{The Luminosity of Compact Nuclear Starbursts and the
Covering Factor of the Dusty Torus}

A compact nuclear starburst can input energy into the torus and
inflate it \citep{fab98}.  
As such an inflated torus has a large covering factor of gas and dust
around the central AGN, the equivalent width of the Fe K$\alpha$ emission
line at 6.4 keV in X-rays (EW$_{\rm Fe}$) can be large.  
The dependence of EW$_{\rm Fe}$ on the covering factor at various
N$_{\rm H}$ has been calculated by several authors
\citep{awa91,ghi94,lev02}.  
Both N$_{\rm H}$ and EW$_{\rm Fe}$ based on
{\it ASCA} data are available for several sources in our sample
\citep{bas99}.   

\citet{lev02} performed this calculation for the Compton-thick case 
(N$_{\rm H}$ $>$ 1 $\times$ 10$^{24}$ cm$^{-2}$). 
Four sources in our sample, NGC 7674, NGC 1667, NGC 4968, and NGC 5135, are
classified as Compton thick and the EW$_{\rm Fe}$s in  NGC 7674, NGC
1667, NGC 4968, and NGC 5135 are estimated to be 0.9, $<$3, 1.18, and
11.7 keV, respectively \citep{bas99}. 
All but NGC 1667 show a detectable compact nuclear starburst in the
torus.
The covering factor of the torus for NGC 7674, NGC 4968, and NGC 5135 is
expected to be $>$40$^{\circ}$ and could be as large as $>$70$^{\circ}$
if N$_{\rm H}$ is close to 1 $\times$ 10$^{24}$ cm$^{-2}$ (Fig. 3 of
Levenson et al. 2002). 
However, the upper limit for EW$_{\rm Fe}$ in NGC 1667, the source with
an undetectable compact nuclear starburst, is so large ($<$3 keV) that 
we are unable to draw meaningful conclusions about the dependence of the
covering factor on the presence or absence of a compact 
nuclear starburst from these four sources.
  
In addition, \citet{ghi94} performed the same calculation for the
Compton-thin case   
(N$_{\rm H}$ $<$ 1 $\times$ 10$^{24}$ cm$^{-2}$).
For the five Compton-thin sources, NGC 4388, NGC 5252, NGC 5674, NGC
262, NGC 7172, both N$_{\rm H}$ and EW$_{\rm Fe}$ are available
\citep{bas99}. 
Compact nuclear starbursts are undetected in all of these sources, but 
the covering factor is expected to be $>$60$^{\circ}$ (Fig. 3 of 
Ghisellini et al. 1994). 

With the datasets currently available, no meaningful constraints can be
obtained for the dependence of the torus covering factor on the 
presence of a compact nuclear starburst. 
However, if more accurate values of N$_{\rm H}$ and EW$_{\rm Fe}$ based
on observations with {\it Chandra} and {\it XMM} (and {\it ASTRO-E2} for
Compton-thick 
sources) become available for a significant number of our sample objects in
the near future, then the dependence can be investigated in much more
detail and the hypothesis that the compact nuclear starburst is the
primary cause of inflation of the dusty molecular torus can be 
tested.   

\section{Summary}

We have presented infrared 2.8--4.1 $\mu$m slit spectra of 32 Seyfert 2
galaxies in the CfA and 12$\mu$m samples, and made the
following main conclusions. 
\begin{enumerate}   
\item By using the 3.3 $\mu$m PAH emission luminosity as an indicator of
      a compact nuclear starburst in the dusty torus, we found evidence
      for the presence of such starbursts in $\sim$30\% of
      Seyfert 2s.
      However, the starbursts generally do not contribute significantly
      to either the observed 3--4 $\mu$m nuclear flux or the total infrared
      luminosity of the observed Seyfert 2 galaxies.   
\item The dust extinction in the direction of AGN-powered 3--4
      $\mu$m-continuum-emitting 
      hot dust in the very inner part of the dusty torus, close to the
      dust sublimation radius, is modest (A$_{\rm V}$ $<$ 50--60 mag) in 
      the majority of the observed Seyfert 2 galaxies.  
\item Provided that the {\it IRAS} 12- and 25-$\mu$m, and nuclear
      $N$-band luminosities are a good measure of AGN power, the AGN 
      and compact-nuclear-starburst luminosities are correlated. 
\end{enumerate}

These conclusions confirm those drawn from a smaller sample of 
Seyfert 2s by \citet{ima02}.

\acknowledgments

We are grateful to Dr. J. Rayner, Dr. H. Terada, Dr. S. Bus, P. Sears,
W. Golisch, B. Potter, D. Scarla, and S. Harasawa for their supports
during our IRTF SpeX and Subaru IRCS observing runs, and to Drs. K. Aoki,
K. Wada, Y. Taniguchi, and M. Elitzur for useful discussions.         
This research has made use of the SIMBAD database, operated at CDS,
Strasbourg, France, and of the NASA/IPAC Extragalactic Database 
(NED) which is operated by the Jet Propulsion Laboratory, California
Institute of Technology, under contract with the National Aeronautics
and Space Administration.

\clearpage

%------------------ Table 1 ------------------%
\begin{deluxetable}{lcrrrrrccc}
\tabletypesize{\scriptsize}
\tablecaption{Infrared Properties of the Observed Seyfert 2
Nuclei. \label{tbl-1}}
\tablenum{1}
\tablewidth{0pt}
\tablehead{
\colhead{Object} & \colhead{Redshift}   & 
\colhead{f$_{\rm N}$}   & 
\colhead{f$_{\rm 12}$}   & 
\colhead{f$_{\rm 25}$}   & 
\colhead{f$_{\rm 60}$}   & 
\colhead{f$_{\rm 100}$}  & 
\colhead{log L$_{\rm IR}$} & 
\colhead{Remarks} \\
\colhead{} & \colhead{} & \colhead{[Jy]} & \colhead{[Jy]} &
\colhead{[Jy]} & \colhead{[Jy]}  & \colhead{[Jy]} & \colhead{[ergs
s$^{-1}$]} & \colhead{} \\ 
\colhead{(1)} & \colhead{(2)} & \colhead{(3)} & \colhead{(4)} & 
\colhead{(5)} & \colhead{(6)} & \colhead{(7)} & \colhead{(8)} &
\colhead{(9)}  
}
\startdata
Mrk 334  & 0.022 & 0.16 $^{a}$ & 0.23 & 1.05 &4.35 & 4.32 & 44.6 & CfA \\
Mrk 993  & 0.015 & 0.018 $^{b}$ & $<$0.13 & $<$0.13 & 0.30 & 1.32 &
43.2--43.6 & CfA \\ 
Mrk 573  & 0.017 & 0.17 $^{b}$ & $<$0.25 & $<$0.93 & 1.09 & 1.26 &
43.6--44.1 & CfA \\ 
NGC 1144 & 0.029 & $<$0.035 $^{c}$ & 0.28 & 0.63 & 5.30 & 11.3 & 45.0
&CfA, 12$\mu$m \\
NGC 4388 & 0.008 & 0.40 $^{d}$ & 1.00 & 3.46 & 10.2 & 18.1 & 44.2 & CfA,
12$\mu$m \\
NGC 5252 & 0.023 & 0.028 $^{a}$ & 0.08 \tablenotemark{A} & 
0.14 \tablenotemark{A} & 1.85 \tablenotemark{A} & 1.13 \tablenotemark{A}
& 44.2 & CfA \\ 
NGC 5256 (Mrk 266SW) & 0.028 & 0.030 $^{a}$ & 0.23 & 0.98 & 7.34 & 11.1
& 45.0 & CfA, 12$\mu$m \\ 
NGC 5347 & 0.008 & 0.21 $^{a}$ & 0.31 & 0.96 & 1.42 & 2.64 & 44.5 & CfA,
12$\mu$m \\
NGC 5674 & 0.025 & 0.024 $^{a}$ & 0.14 & 0.28 & 1.44 & 3.73 & 44.4 & CfA \\
NGC 5695 & 0.014 & $<$0.01 $^{a}$ & $<$0.11 & 0.13 & 0.57 & 1.79 &
43.4--43.6 & CfA \\ 
NGC 5929 & 0.008 & 0.024 $^{a}$ & 0.43 & 1.62 & 9.14 & 13.7 & 44.1 & CfA,
12$\mu$m \\
NGC 7674 & 0.029 & 0.33 $^{a}$ & 0.67 & 1.90 & 5.59 & 8.15 & 45.1 & CfA,
12$\mu$m \\
NGC 7682 & 0.017 & 0.012 $^{a}$ & 0.32 \tablenotemark{A} & 0.22
\tablenotemark{A} & 0.47 \tablenotemark{A} & 0.41 \tablenotemark{A} &
43.9 & CfA \\  \hline
Mrk 938  & 0.019 & 0.24 $^{a}$ & 0.40 & 2.37 & 16.6 & 17.2 & 45.0
& 12$\mu$m \\
NGC 262 (Mrk 348) & 0.015 & 0.30 $^{e}$ & 0.31 & 0.83 & 1.29 & 1.55 &
44.0 & 12$\mu$m \\
NGC 513  & 0.020 & & 0.17 & 0.28 & 1.94 & 4.05 & 44.3 & 12$\mu$m \\
F01475$-$0740 & 0.017 & 0.21 $^{f}$ & 0.31 & $<$0.96 & 1.05 & 0.65 &
43.9--44.1 & 12$\mu$m \\ 
NGC 1125 & 0.011 & & 0.17 & 0.83 & 3.31 & 3.94 & 43.9 & 12$\mu$m \\
NGC 1194 & 0.013 & & 0.27 & 0.51 & 0.77 & $<$0.93 & 43.7 & 12$\mu$m \\
NGC 1241 & 0.014 & 0.08 $^{a}$ & 0.24 & 0.45 & 3.56 & 10.3 & 44.2
& 12$\mu$m \\
NGC 1320 (Mrk 607) & 0.010 & 0.028$^{a}$ & 0.33 & 1.07 & 2.15 & 2.75 &
43.8 & 12$\mu$m \\
F04385$-$0828 & 0.015 & 0.27 $^{a}$ & 0.42 & 1.62 & 2.77 & 2.53 & 44.3
& 12$\mu$m \\
NGC 1667 & 0.015 & 0.005 $^{a}$ & 0.43 & 0.68 & 5.95 & 14.7 & 44.5
& 12$\mu$m \\
NGC 3660 & 0.012 & 0.033 $^{a}$ & 0.19 & 0.22 & 1.87 & 4.54 & 43.8
& 12$\mu$m \\
NGC 4501 & 0.008 & 0.006 $^{a}$ & 1.02 & 1.28 & 13.7 & 54.7 & 44.4
& 12$\mu$m \\
NGC 4968 & 0.010 & 0.12 $^{a}$ & 0.39 & 1.05 & 2.38 & 2.94 & 43.8
& 12$\mu$m \\
MCG-3-34-64 & 0.017 & 0.44 $^{a}$ & 0.88 & 2.86 & 5.90 & 5.48 & 44.7
& 12$\mu$m  \\
NGC 5135 & 0.014 & 0.16 $^{a}$ & 0.64 & 2.40 & 16.9 & 28.6 & 44.8
& 12$\mu$m \\
MCG-2-40-4 & 0.024 & & 0.39 & 0.86 & 3.65 & 6.53 & 44.7 & 12$\mu$m \\
F15480$-$0344 & 0.030 & & 0.18 & 0.73 & 1.07 & $<$4.05 & 44.5--44.6
& 12$\mu$m \\
NGC 7172 & 0.009 & & 0.44 & 0.76 & 5.71 & 12.3 & 44.0 & 12$\mu$m \\
MCG-3-58-7 & 0.032 & & 0.28 & 0.80 & 2.42 & 3.36 & 44.8 & 12$\mu$m \\
\enddata

\tablecomments{
Column (1): Object. CfA Seyfert 2s are placed first. Ordered by right
ascension in each Seyfert 2 sample. 
Column (2): Redshift.
Column (3): Ground-based $N$-band (10.6 $\mu$m) photometry and
references. (a): \citet{mai95}, (b): \citet{ede87}, (c): \citet{car88}, 
(d): \citet{dev87}, (e): \citet{rie78}, (f): \citet{hil88}. 
Columns (4)--(7): f$_{12}$, f$_{25}$, f$_{60}$, and f$_{100}$ are 
{\it IRAS FSC}
fluxes at 12$\mu$m, 25$\mu$m, 60$\mu$m, and 100$\mu$m, respectively.
Column (8): Logarithm of infrared (8$-$1000 $\mu$m) luminosity
in ergs s$^{-1}$ calculated with
$L_{\rm IR} = 2.1 \times 10^{39} \times$ D(Mpc)$^{2}$
$\times$ (13.48 $\times$ $f_{12}$ + 5.16 $\times$ $f_{25}$ +
$2.58 \times f_{60} + f_{100}$) ergs s$^{-1}$
\citep{san96}.
Column (9): CfA \citep{huc92} or 12 $\mu$m \citep{rus93} Seyfert 2s.
}

\tablenotetext{A}{ISO measurements \citep{per01}. The f$_{\rm 12}$ and
f$_{\rm 100}$ are extrapolated and interpolated, respectively, from
measurements at other wavelengths.}

\end{deluxetable}

\clearpage

%-------------- Table 2 ----------------%
\begin{deluxetable}{llcclccc}
\tabletypesize{\scriptsize}
\tablecaption{Observing Log. \label{tbl-2}}
\tablenum{2}
\tablewidth{0pt}
\tablehead{
\colhead{} & \colhead{Date} & \colhead{Telescope \&} &
\colhead{Integration} & \multicolumn{4}{c}{Standard Star} \\
\cline{5-8} \\
\colhead{Object} & \colhead{(UT)} & \colhead{Instrument} & 
\colhead{Time (min)} & \colhead{Star Name} & \colhead{$L$-mag}  &
\colhead{Type} & \colhead{T$_{\rm eff}$ (K)} 
}
\startdata
Mrk 334  & 2002 Aug 19 & Subaru IRCS & 20 & HR 8955 & 5.1 & F6V & 6400 \\
Mrk 993  & 2002 Aug 27, 29 & IRTF SpeX & 100 & HR 410 & 5.0 & F7V & 6240 \\
Mrk 573  & 2002 Aug 28 & IRTF SpeX & 30 & HR 650 & 4.1 & F8V & 6000 \\
NGC 1144 & 2002 Oct 24 & Subaru IRCS & 16 & HR 962 & 3.7 & F8V & 6000 \\
NGC 4388 & 2003 Mar 18 & IRTF SpeX & 30 & HR 4708 & 5.0 & F8V & 6000 \\
NGC 5252 & 2003 Mar 19 & IRTF SpeX & 40 & HR 5011 & 3.8  & G0V & 5930 \\ 
NGC 5256 & 2001 Apr 9 & IRTF NSFCAM & 36 & HR 4767 & 4.8 & F8V--G0V & 6000 \\
NGC 5347 & 2003 Mar 18 & IRTF SpeX & 40 & HR 5346 & 4.8 & F8V & 6000 \\
NGC 5674 & 2002 Mar 28 & Subaru IRCS & 12 & HR 5386 & 5.1 & A0V & 9480  \\
NGC 5695 & 2003 Mar 20 & IRTF SpeX & 60 & HR 5630 & 4.9 & F8V & 6000 \\
NGC 5929 & 2002 Aug 29 & IRTF SpeX \tablenotemark{a} & 40 & HR 5728 &
4.5 & G3V & 5800 \\ 
NGC 7674 & 2002 Aug 27 & IRTF SpeX & 20 & HR 8653 & 4.6 & G8IV & 5400 \\
NGC 7682 & 2002 Aug 27, 29 & IRTF SpeX & 160 & HR 8969 & 2.8 & F7V &
6240 \\  
\hline
Mrk 938  & 2002 Aug 28 & IRTF SpeX & 30 & HR 8917 \tablenotemark{b} &
4.9 & G0V & 5930 \\
NGC 262  & 2002 Aug 28 & IRTF SpeX & 30 & HR 410 & 5.0 & F7V & 6240 \\
NGC 513  & 2002 Aug 28 & IRTF SpeX & 30 & HR 410 & 5.0 & F7V & 6240  \\
F01475$-$0740 & 2002 Aug 27 & IRTF SpeX & 30 & HR 466 & 4.9 & F7V & 6240 \\
NGC 1125 & 2002 Oct 24 & Subaru IRCS & 8 & HR 784 & 4.5 & F6V & 6400 \\
NGC 1194 & 2002 Aug 28 & IRTF SpeX & 30 & HR 996 & 3.2 & G5V & 5700 \\
NGC 1241 & 2002 Aug 29 & IRTF SpeX & 50 & HR 784 & 4.5 & F6V & 6400 \\
NGC 1320 & 2002 Aug 29 & IRTF SpeX & 30 & HR 784 & 4.5 & F6V & 6400 \\
F04385$-$0828 & 2002 Aug 29 & IRTF SpeX & 30 & HR 1536 & 4.4 & F8V & 6000 \\
NGC 1667 & 2003 Mar 20 & IRTF SpeX & 60 & HR 1536 & 4.4 & F8V & 6000 \\
NGC 3660 & 2003 Mar 20 & IRTF SpeX & 90 & HR 4529 & 4.9 & F7V & 6240 \\
NGC 4501 & 2003 Mar 18 & IRTF SpeX & 40 & HR 4708 & 5.0 & F8V & 6000 \\
NGC 4968 & 2003 Mar 20 & IRTF SpeX & 30 & HR 4935 & 4.2 & F7V & 6240 \\
MCG-3-34-64 & 2003 Mar 19 & IRTF SpeX & 40 & HR 4995 & 3.6 & G6V & 5620 \\
NGC 5135 & 2001 Apr 8 & IRTF NSFCAM & 36 & HR 5212 & 4.8 & F7V & 6240\\
MCG-2-40-4 & 2003 Mar 19 & IRTF SpeX & 40 & HR 5779 & 5.2 & F7V & 6240 \\
F15480$-$0344 & 2002 Aug 29 & IRTF SpeX & 30 & HR 5779 & 5.2 & F7V & 6240 \\
NGC 7172 & 1999 Sep 9 & UKIRT CGS4 & 53.3 & HR 8087 & 5.3 & A0V & 9480 \\
MCG-3-58-7 & 2002 Aug 28 & IRTF SpeX & 20 & HR 8457 & 4.8 & F6V & 6400 \\
\enddata

\tablenotetext{a}{3--4 $\mu$m spectrum taken with IRTF NSFCAM was
presented in \citet{ima02}, but the presence of the 3.3 $\mu$m PAH
emission was unclear. This source was re-observed using IRTF SpeX.}

\tablenotetext{b}{Flux calibration was made using HR 8457.}

\end{deluxetable}

\clearpage

%--------------- Table 3 ----------------%
\begin{deluxetable}{lcccc}
\tabletypesize{\scriptsize}
\tablecaption{The Properties of the 3.3 $\mu$m PAH Emission
Feature. \label{tbl-3}} 
\tablenum{3}
\tablewidth{0pt}
\tablehead{
\colhead{} & \colhead{f$_{3.3 \rm PAH}$} & \colhead{L$_{3.3 \rm PAH}$}  &
\colhead{L$_{3.3 \rm PAH}$/L$_{\rm IR}$} & \colhead{rest EW$_{3.3 \rm PAH}$} \\
\colhead{Object} & \colhead{($\times$ 10$^{-14}$ ergs s$^{-1}$ cm$^{-2}$)} & 
\colhead{($\times$ 10$^{39}$ ergs s$^{-1}$)} & 
\colhead{($\times$ 10$^{-3}$)} 
& \colhead{(nm)} \\
\colhead{(1)} & \colhead{(2)} & \colhead{(3)} & \colhead{(4)} 
& \colhead{(5)}  
}
\startdata
Mrk 334  & 6.1$\pm$0.1 & 58$\pm$1 & 0.1     & 11  \\
Mrk 993  & $<$2.2      & $<$9.4   & $<$0.6  & $<$16 \\
Mrk 573  & $<$3.2      & $<$18    & $<$0.5  & $<$6 \\
NGC 1144 & $<$0.7      & $<$12    & $<$0.02 & $<$7 \\
NGC 4388 & $<$3.8  & $<$4.7 & $<$0.03 & $<$5\\
NGC 5252 & $<$5.4 & $<$56.4 & $<$0.4 & $<$12 \\
NGC 5256 & 8.5$\pm$0.5 & 133$\pm$7 & 0.1 \tablenotemark{a} & 87 \\
NGC 5347 & $<$6.4 & $<$7.9 & $<$0.3 & $<$10 \\
NGC 5674 & $<$1.5      & $<$19     & $<$0.08 & $<$4 \\
NGC 5695 & $<$1.6 & $<$6.1 & $<$0.3 & $<$16 \\
NGC 5929 & $<$1.4      & $<$1.7    & $<$0.02 & $<$13 \\
NGC 7674 & 8.1$\pm$0.5 & 136$\pm$8 & 0.1     & 7 \\
NGC 7682 & 2.0$\pm$0.5 & 11$\pm$3 & 0.2 & 40 \\ 
\hline
Mrk 938  & 41$\pm$1 & 289$\pm$3 & 0.3    & 75   \\
NGC 262  & $<$8.6   & $<$38     & $<$0.4 & $<$7 \\
NGC 513  & $<$2.7   & $<$22     & $<$0.2 & $<$6 \\
F01475-0740 & 3.9$\pm$0.5 & 22$\pm$3 & 0.3 & 21 \\
NGC 1125 & 10.2$\pm$1.4 & 24$\pm$3 & 0.3 & 50 \\
NGC 1194 & $<$2.7 & $<$8.8 & $<$0.2 & $<$4 \\
NGC 1241 & $<$1.1 & $<$4.1 & $<$0.03 & $<$12 \\
NGC 1320 & $<$5.9 & $<$12 & $<$0.2 & $<$8 \\
F04385$-$0828 & 4.1$\pm$0.3 & 18$\pm$1 & 0.1 & 4 \\
NGC 1667 & $<$1.6 & $<$7.1 & $<$0.03 & $<$16 \\
NGC 3660 & 4.0$\pm$1.1 & 11$\pm$3 & 0.2 & 50 \\
NGC 4501 & $<$3.8 & $<$4.7 & $<$0.02 & $<$12 \\
NGC 4968 & 8.5$\pm$1.1 & 17$\pm$2 & 0.2 & 18 \\
MCG-3-34-64 & $<$3.3 & $<$18 & $<$0.04 & $<$6 \\
NGC 5135 & 15$\pm$3 & 58$\pm$11 & 0.09 & 24 \\
MCG-2-40-4 & $<$7.1 & $<$80 & $<$0.2 & $<$3 \\
F15480$-$0344 & $<$3.3 & $<$59 & $<$0.2 & $<$11 \\
NGC 7172   & $<$5.3 \tablenotemark{b} & $<$8.4 \tablenotemark{b} & 
$<$0.08 \tablenotemark{b} & $<$3.5 \tablenotemark{b} \\
MCG-3-58-7 & $<$3.8 & $<$78  & $<$0.2 & $<$3 \\
\enddata

\tablecomments{Column (1): Object.
Column (2): Observed 3.3 $\mu$m PAH flux.
Column (3): Observed 3.3 $\mu$m PAH luminosity.
Column (4): Observed 3.3 $\mu$m PAH to infrared luminosity ratio
            in units of 10$^{-3}$, the typical value for starburst
            galaxies.
Column (5): Rest frame equivalent width of the 3.3 $\mu$m PAH emission.}

\tablenotetext{a}{NGC 5256 is a double-nuclei source. 
The L$_{\rm 3.3PAH}$ is for only one nucleus (Mrk 266 SW), whereas both
nuclei contribute to L$_{\rm IR}$.} 

\tablenotetext{b}{An upper limit was re-calculated after \citet{ima00}.}

\end{deluxetable}

\clearpage

%--------------  Figure 1 -------------%

\begin{figure}
\plottwo{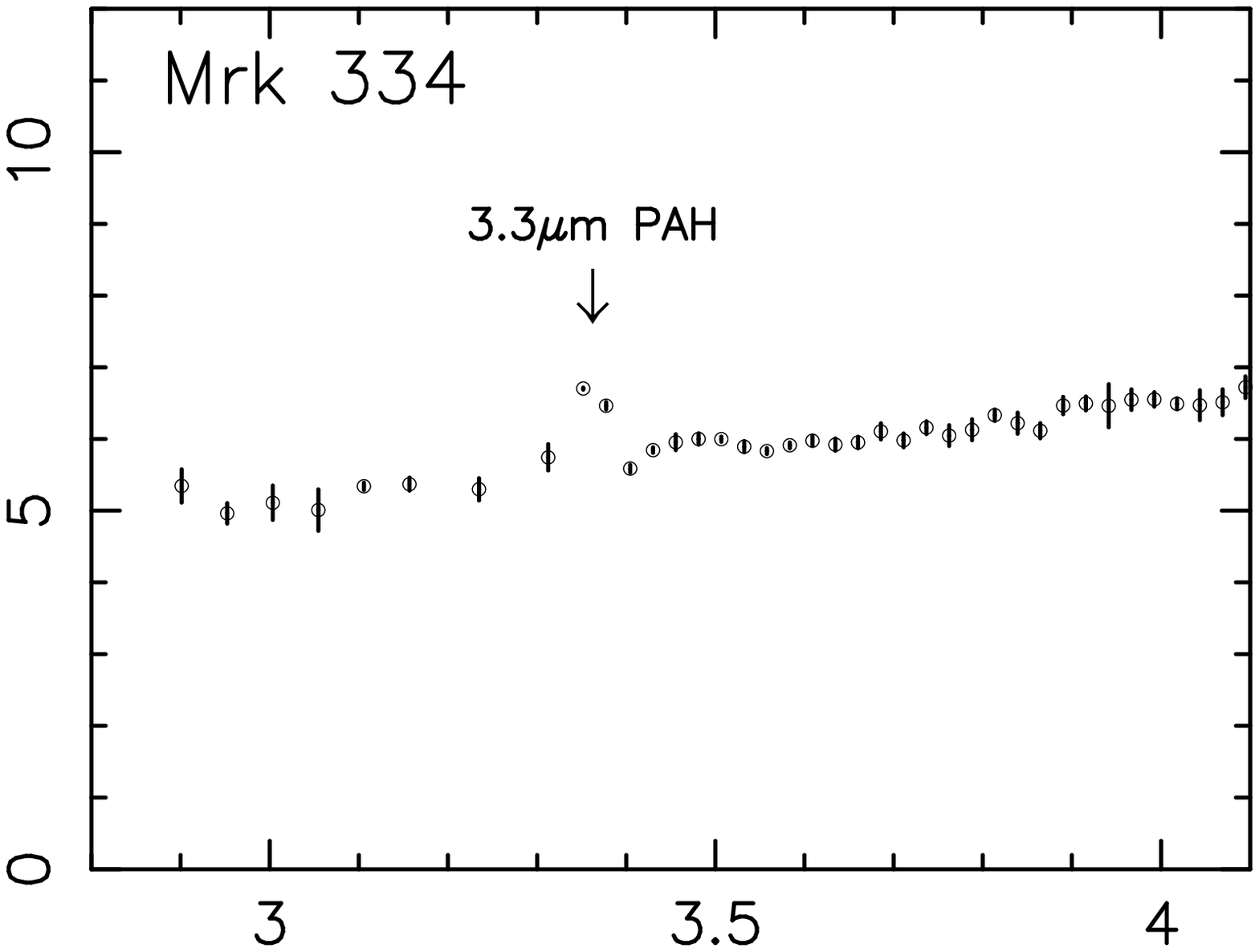}{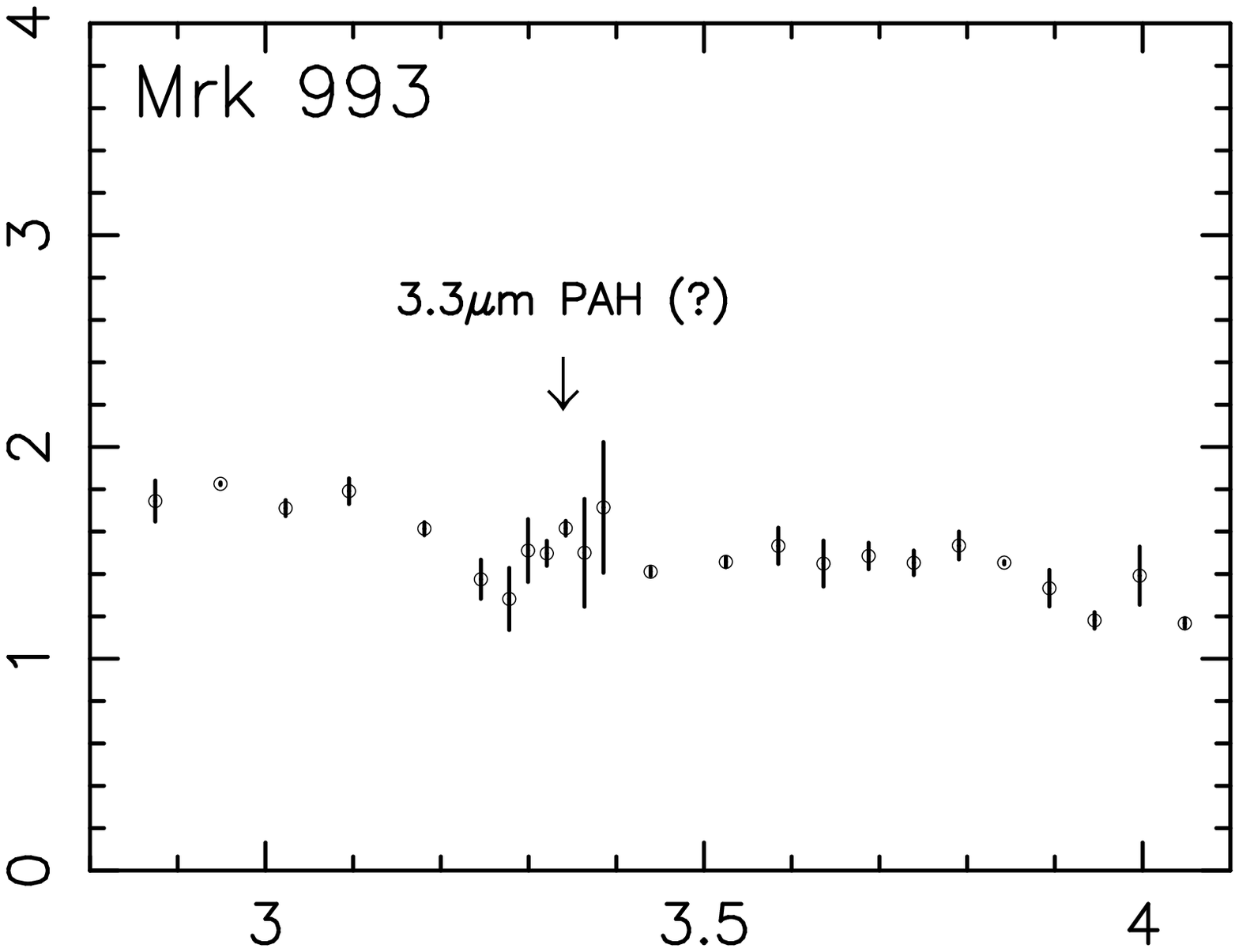} 
\end{figure}
\begin{figure} 
\plottwo{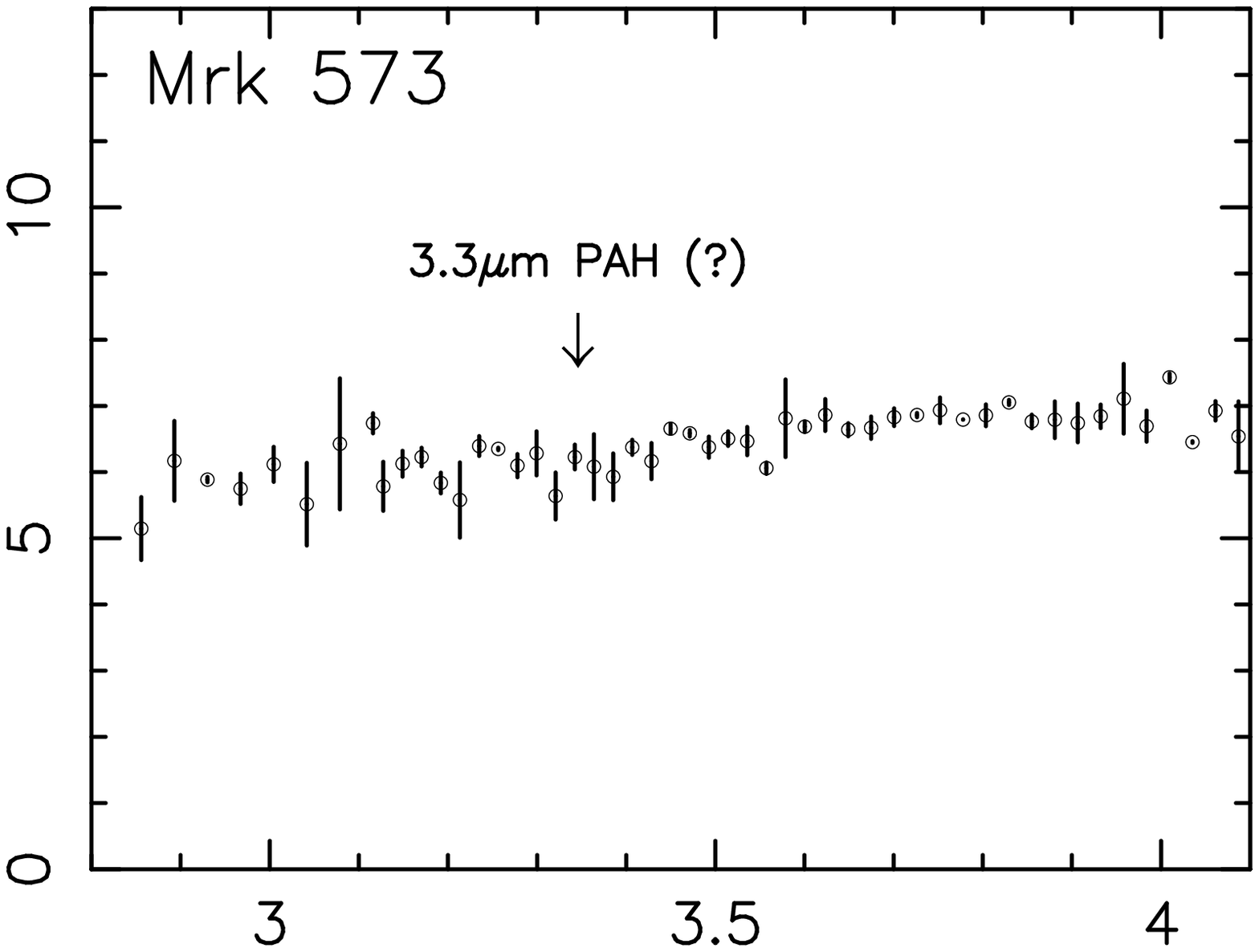}{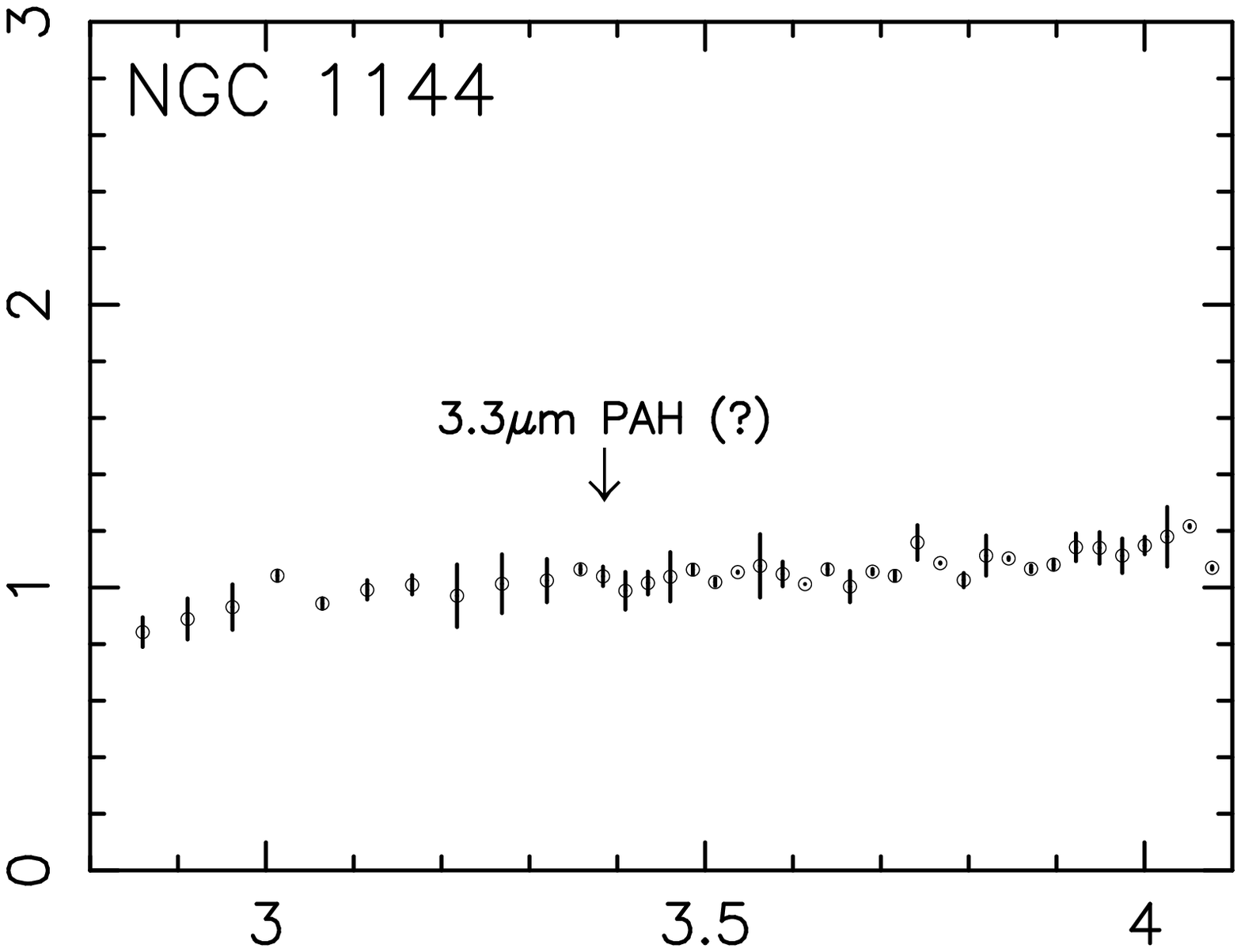}
\end{figure}
\begin{figure} 
\plottwo{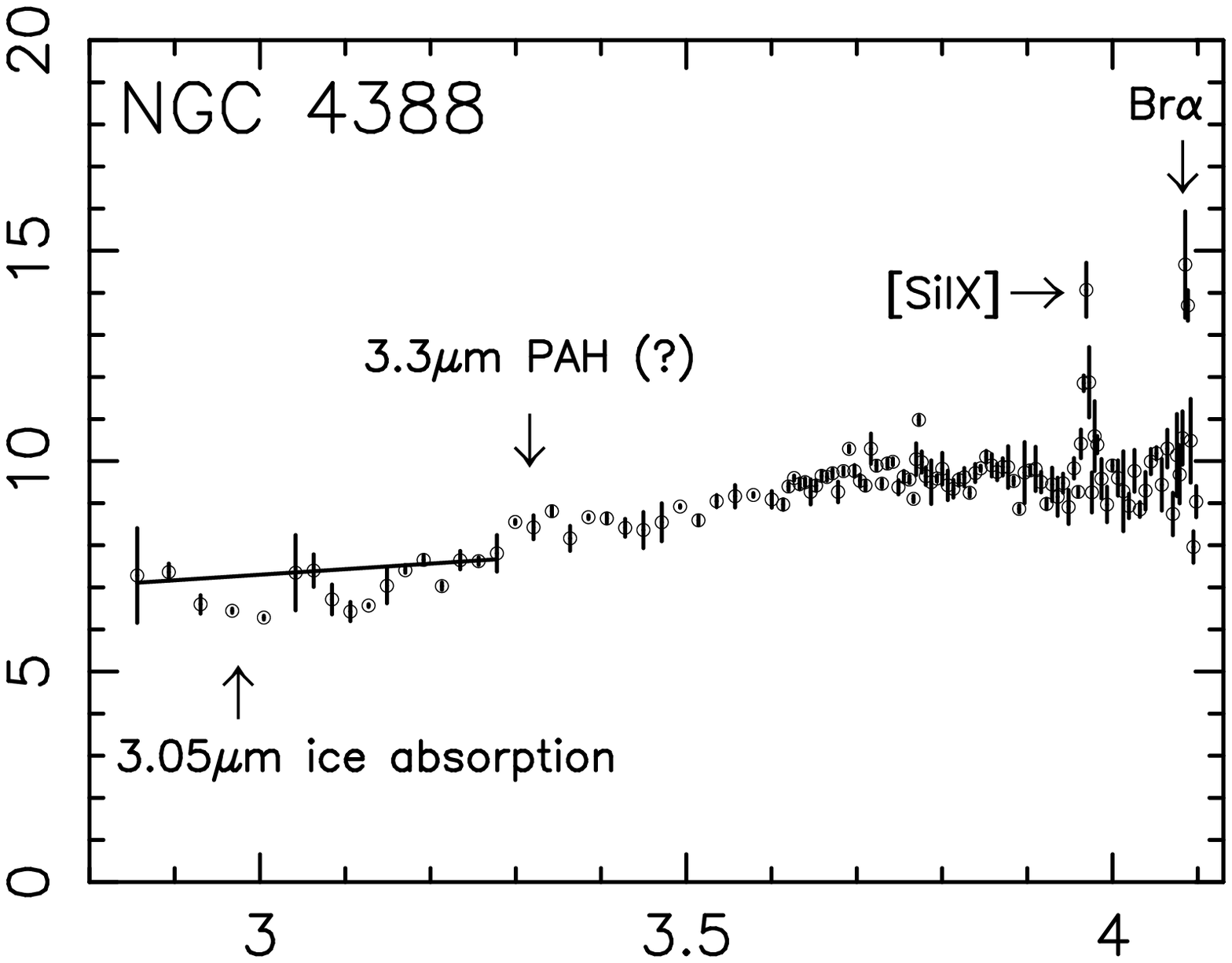}{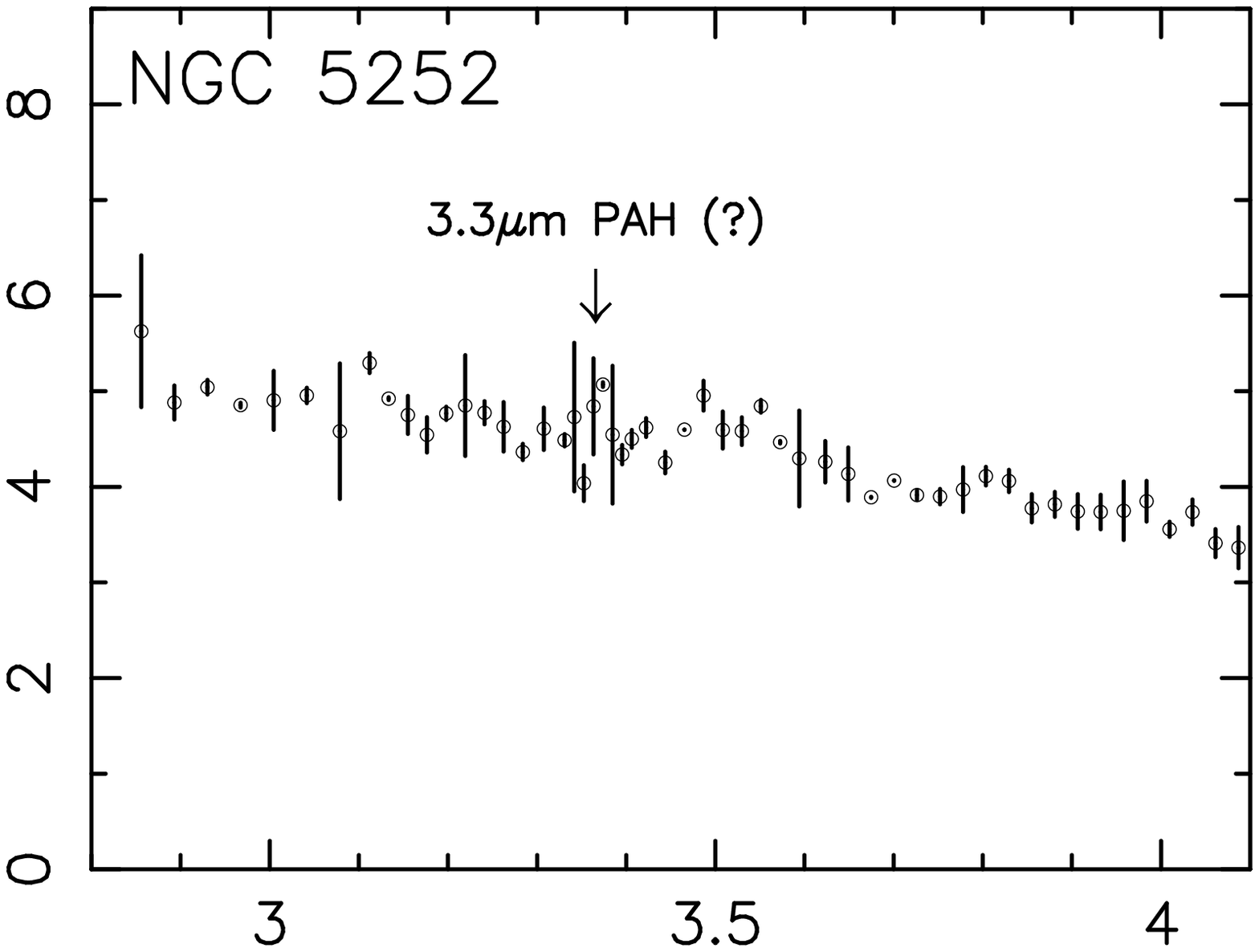}
\end{figure}

\clearpage

\begin{figure}
\plottwo{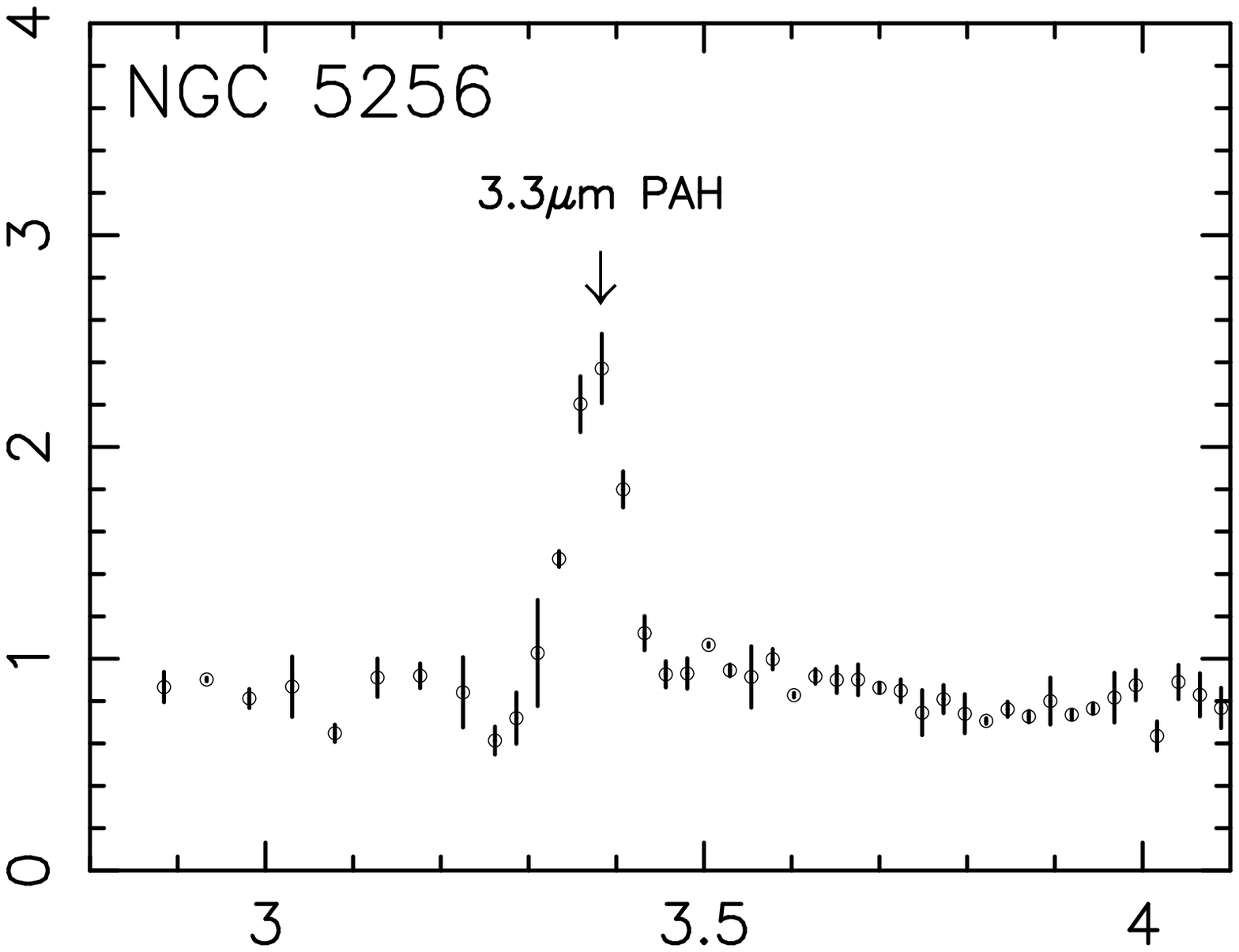}{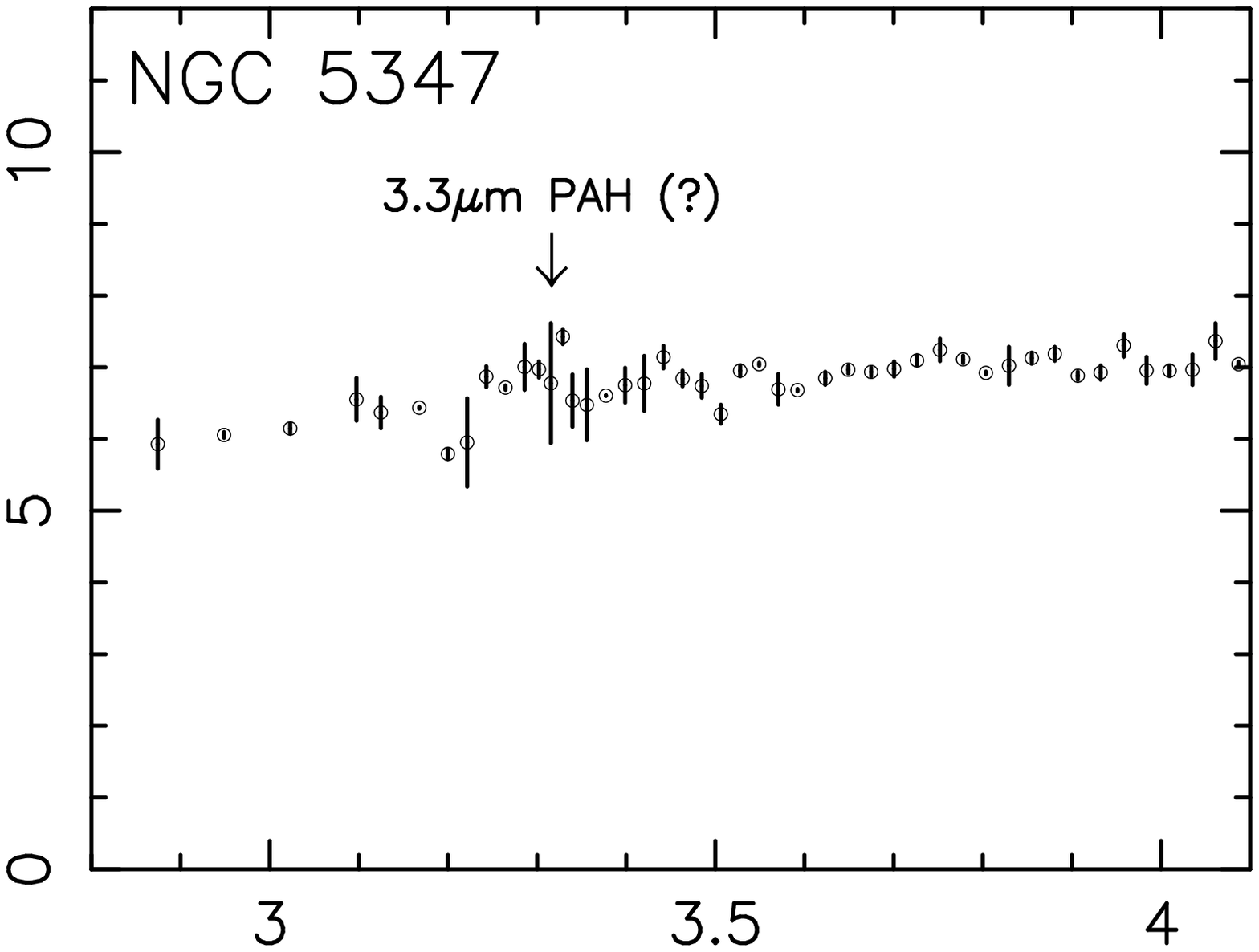}
\end{figure}
\begin{figure}
\plottwo{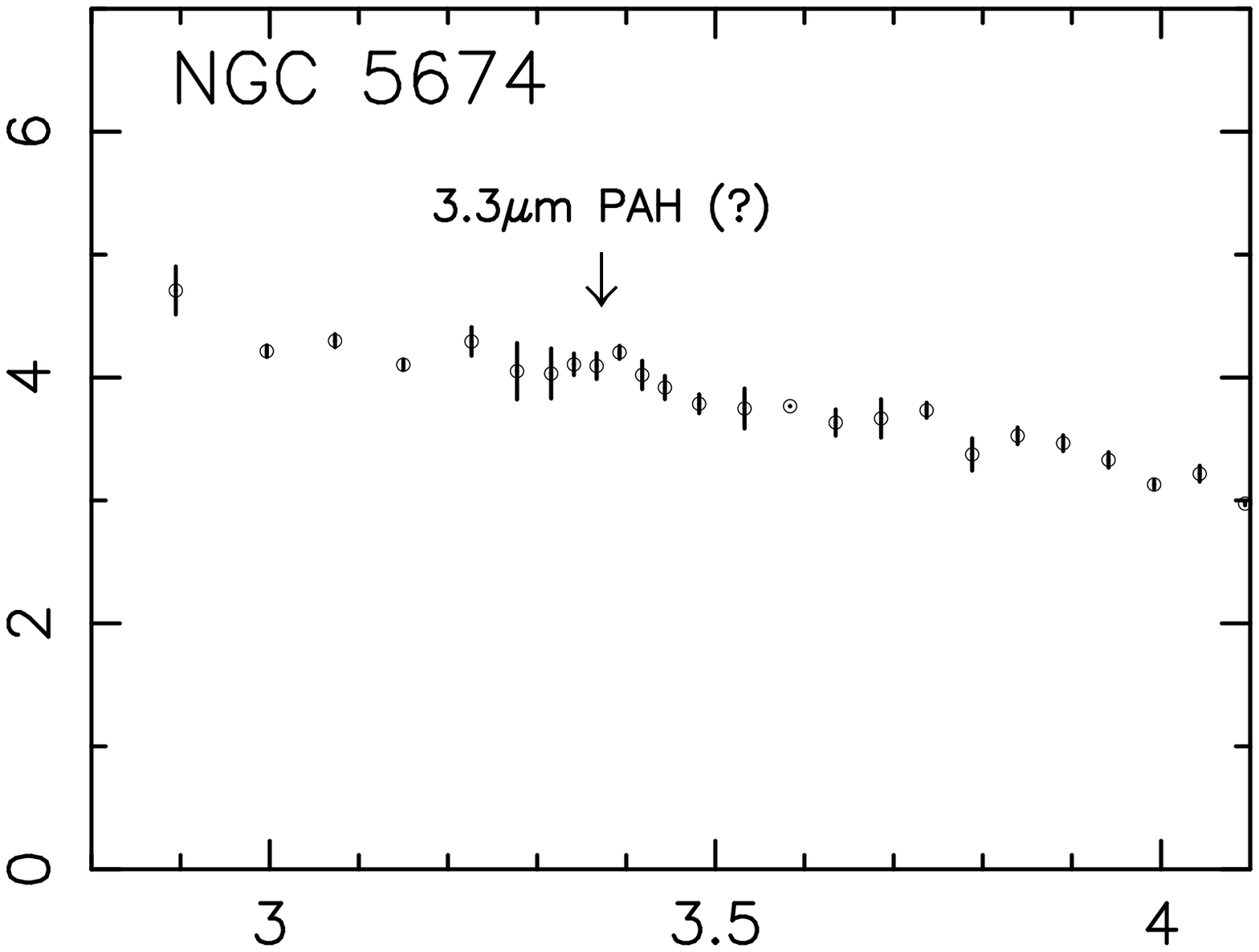}{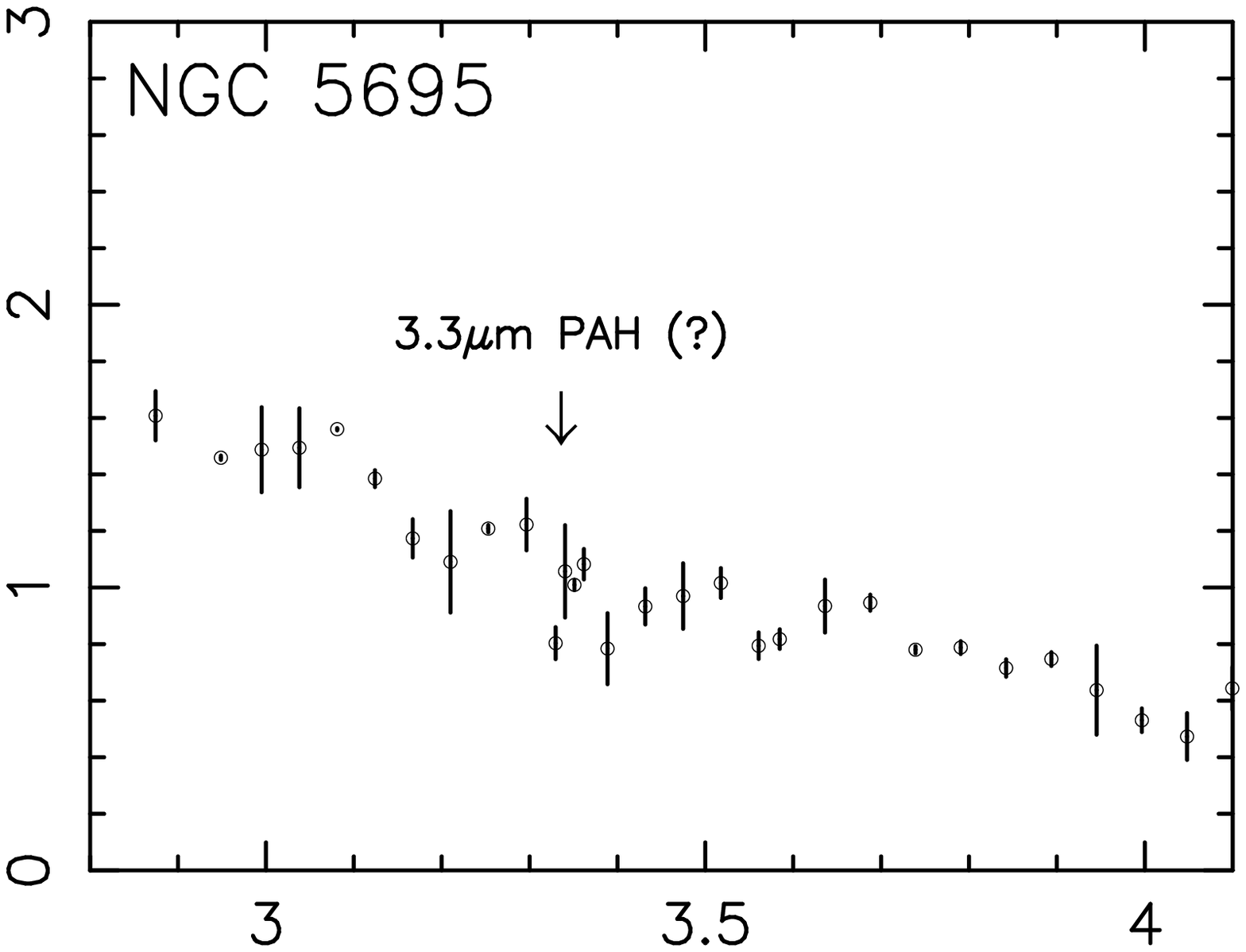}
\end{figure}
\begin{figure} 
\plottwo{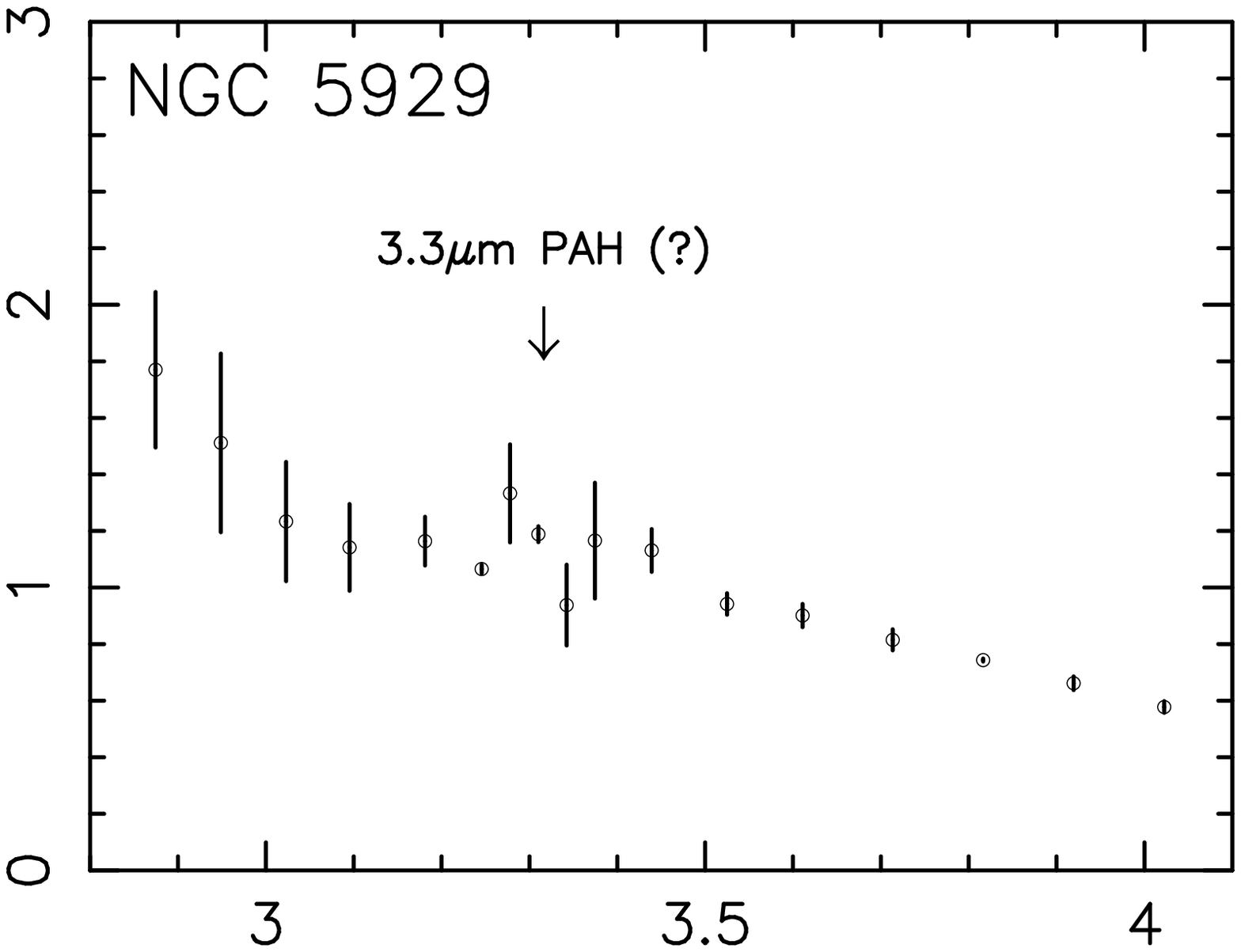}{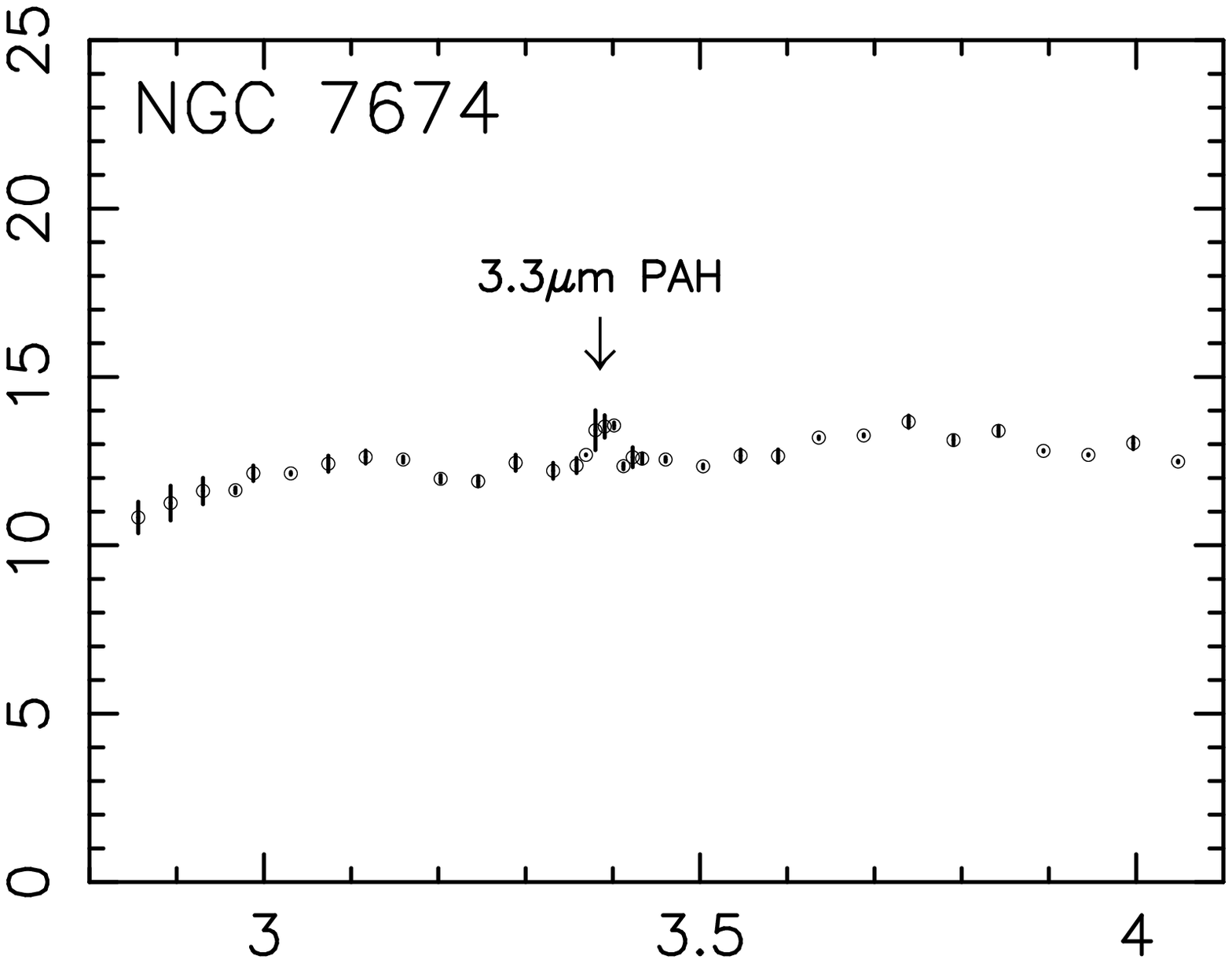}
\end{figure}

\clearpage

\begin{figure} 
\plottwo{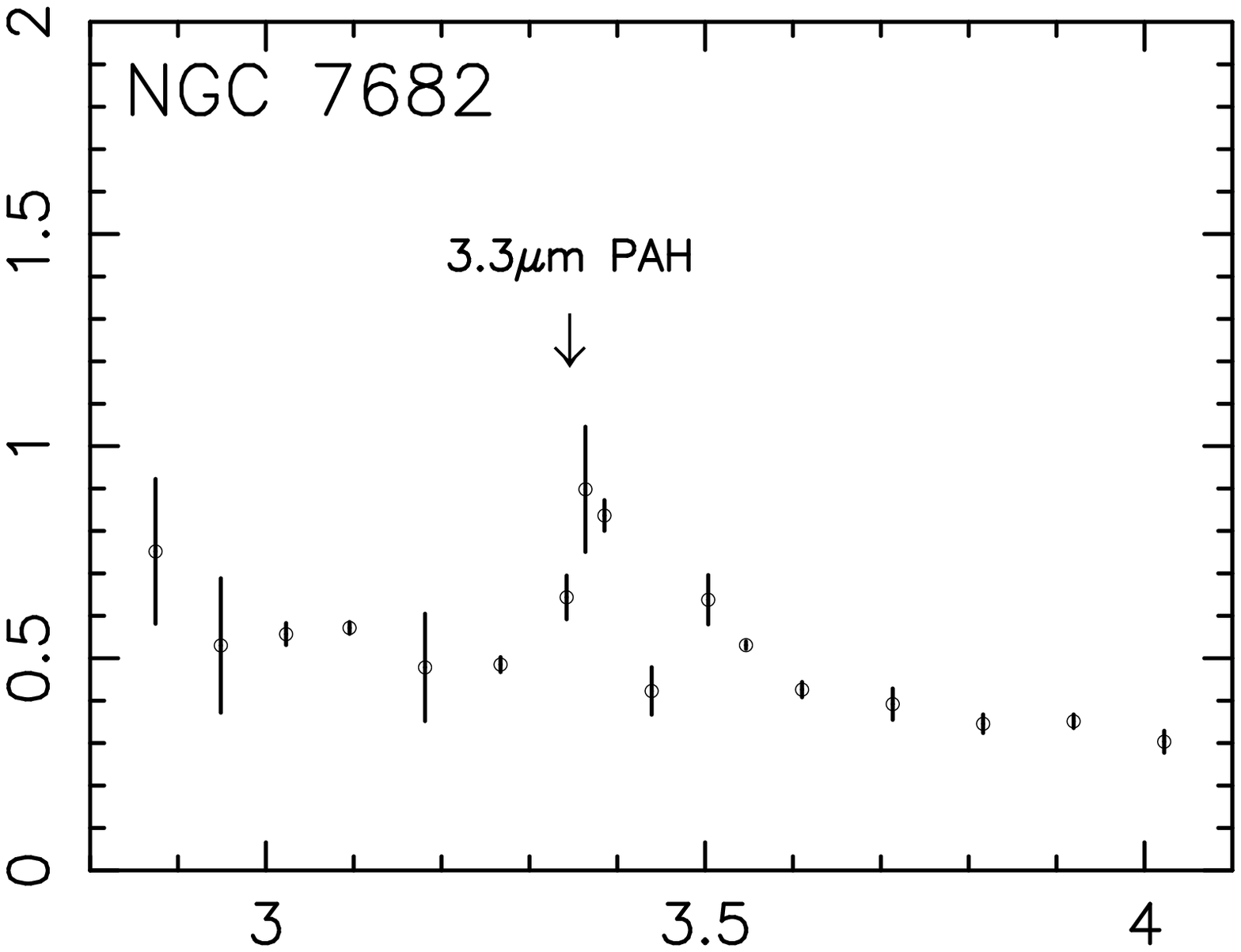}{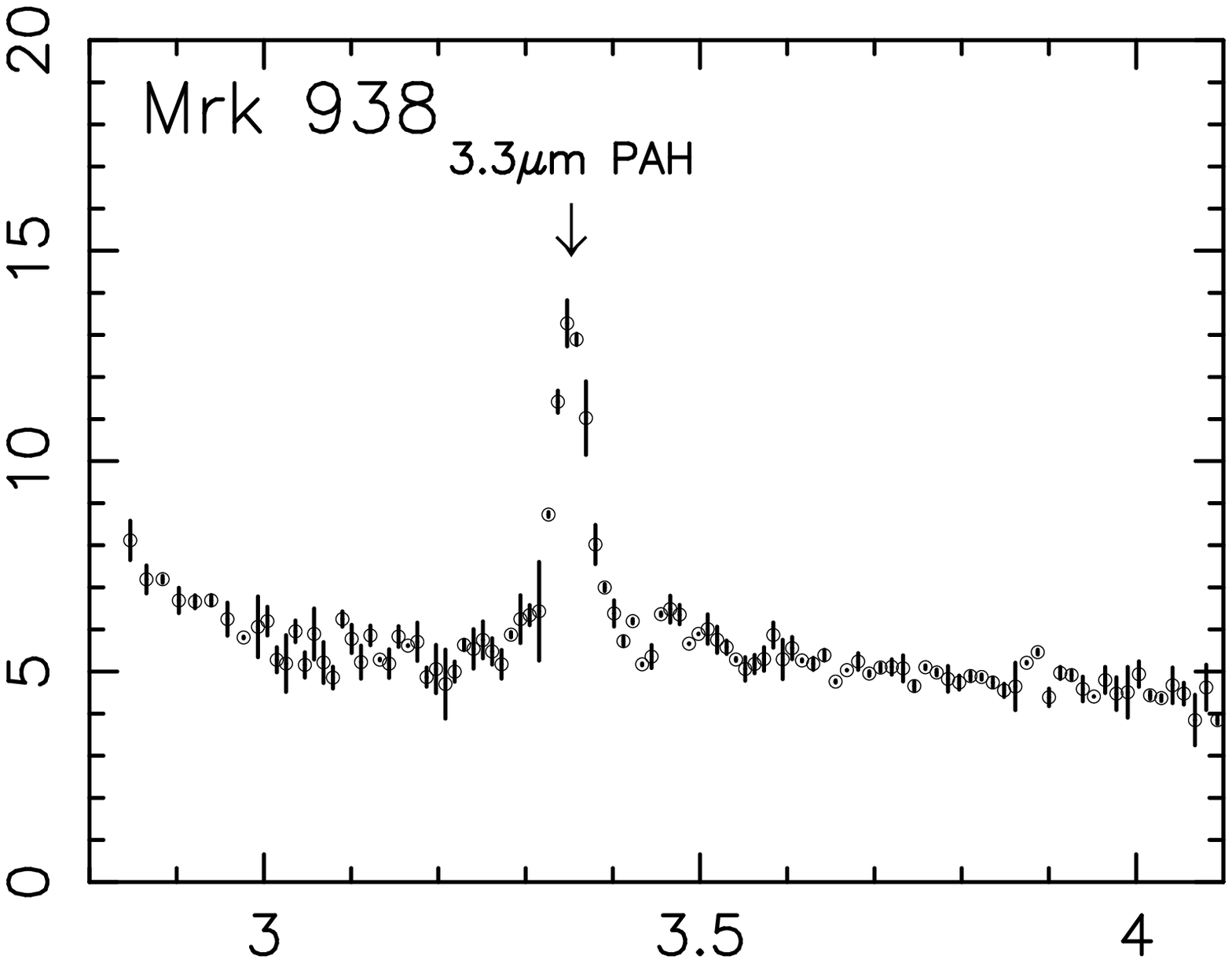}
\end{figure}
\begin{figure} 
\plottwo{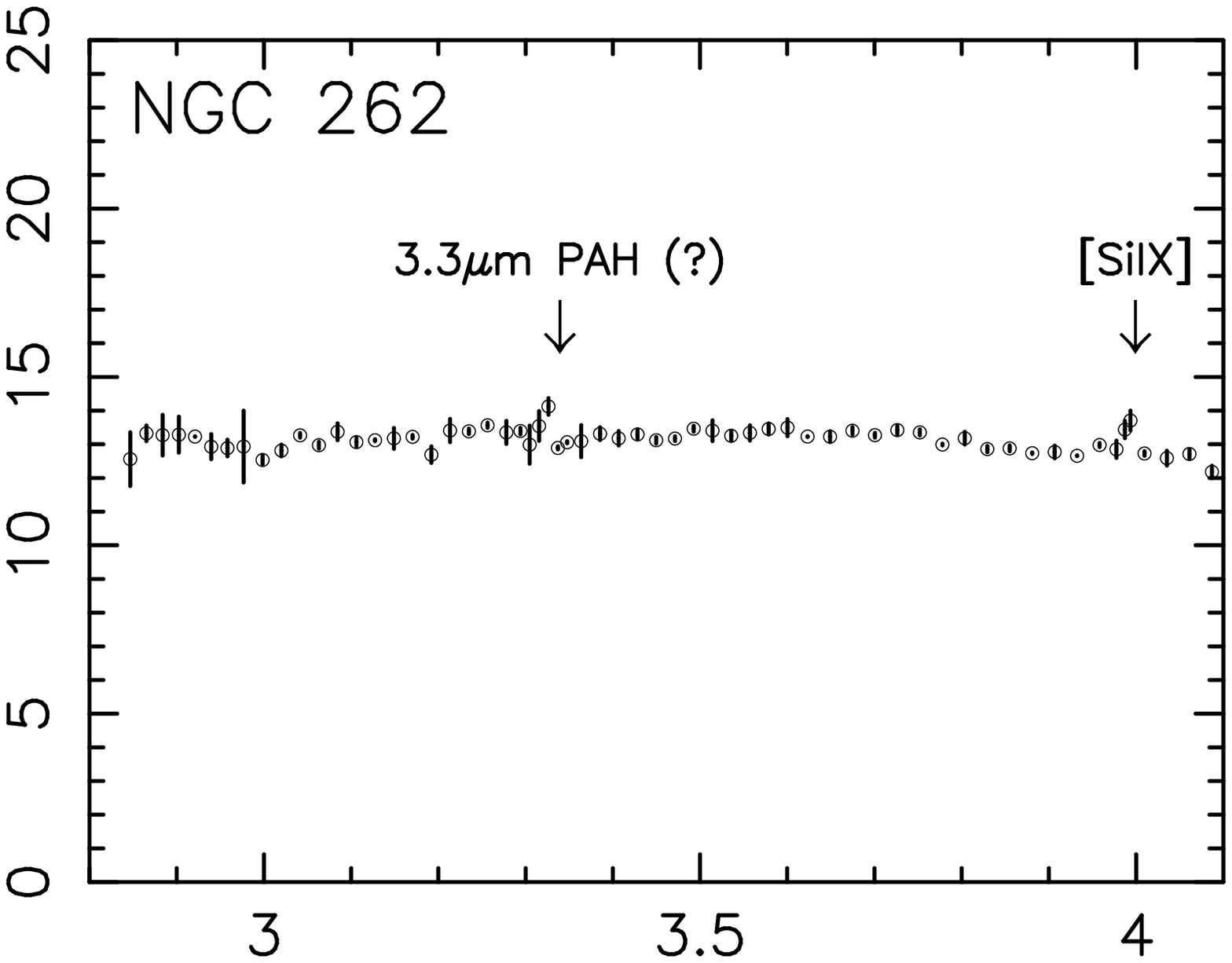}{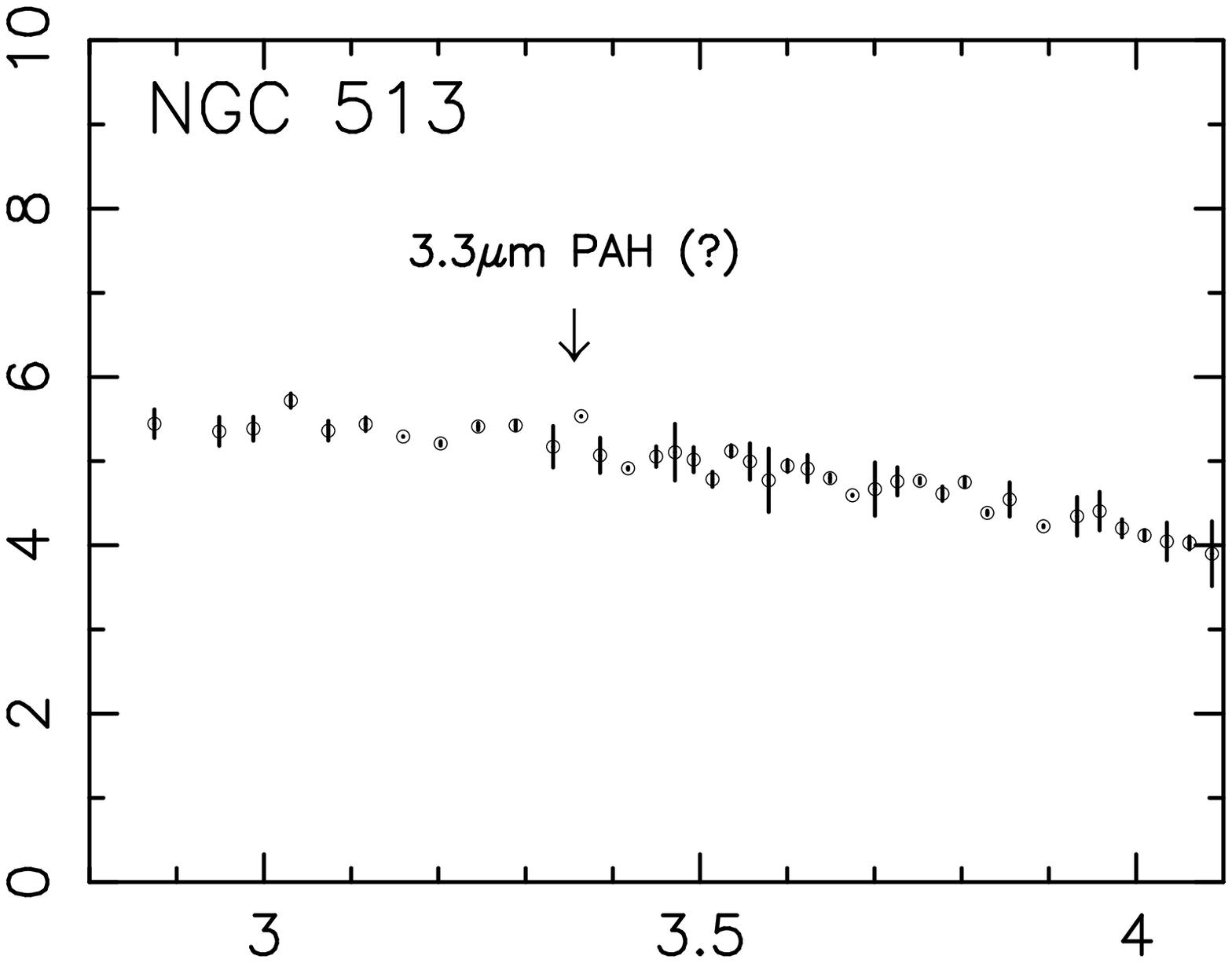}
\end{figure}
\begin{figure}
\plottwo{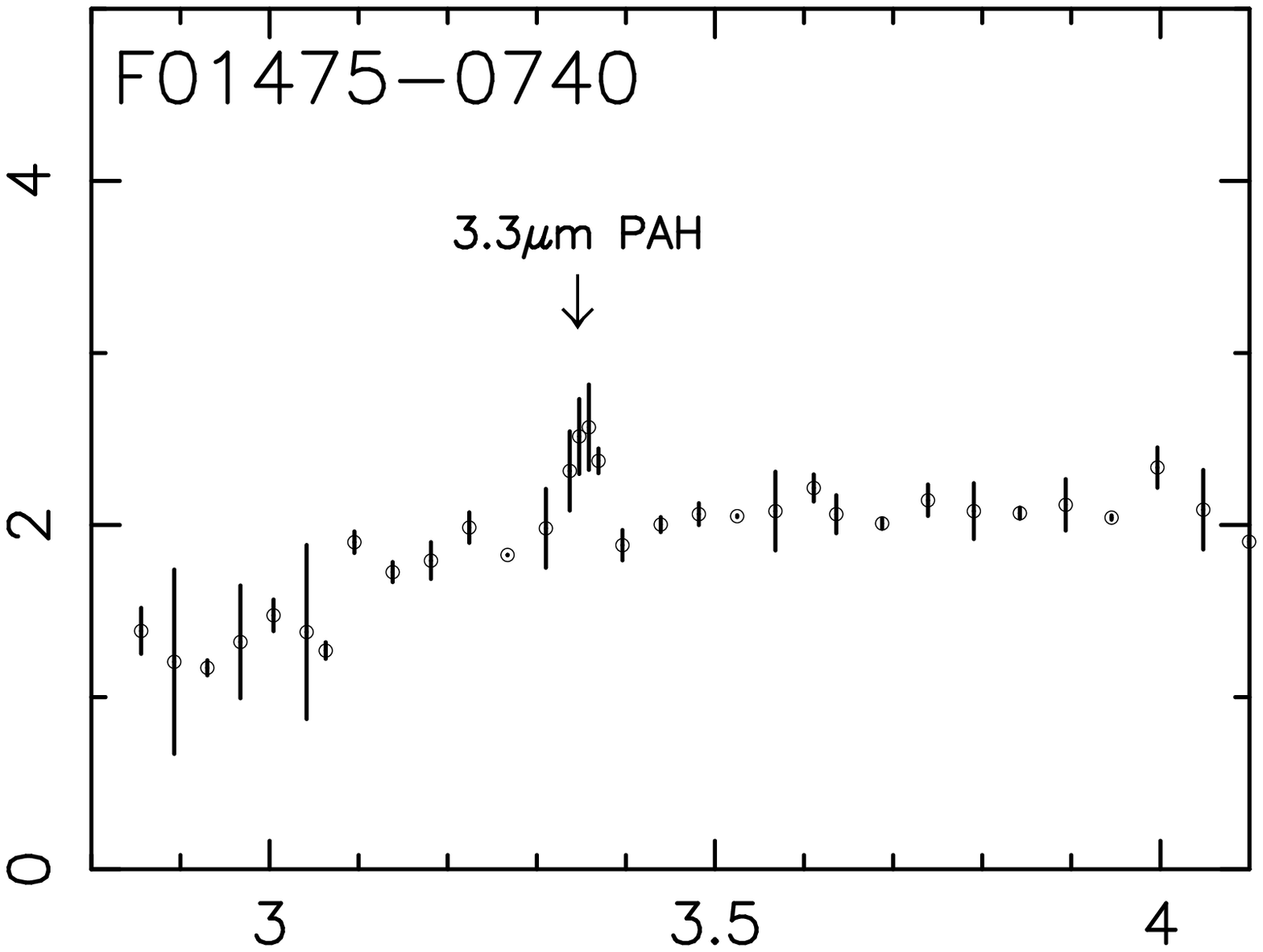}{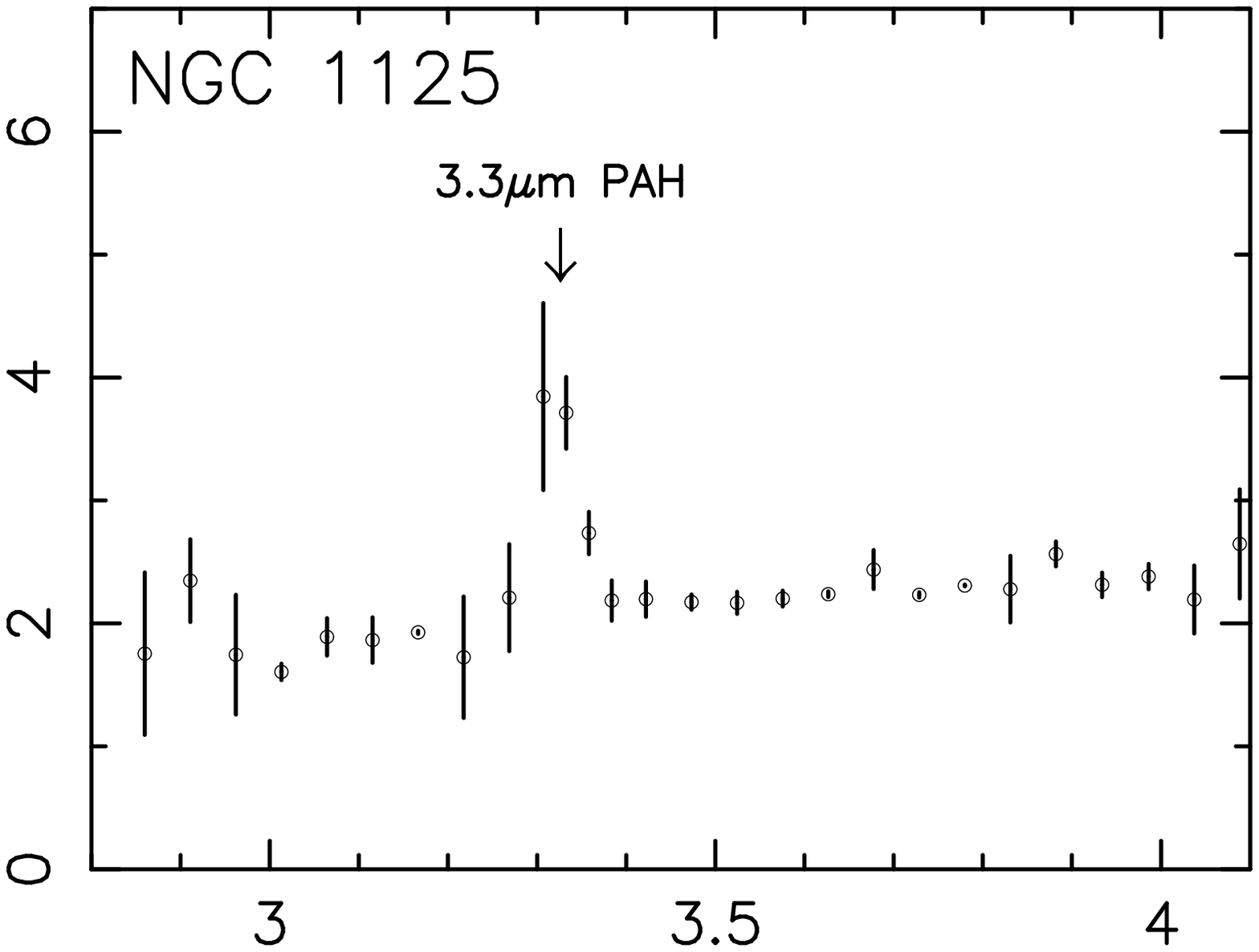}
\end{figure}

\clearpage

\begin{figure} 
\plottwo{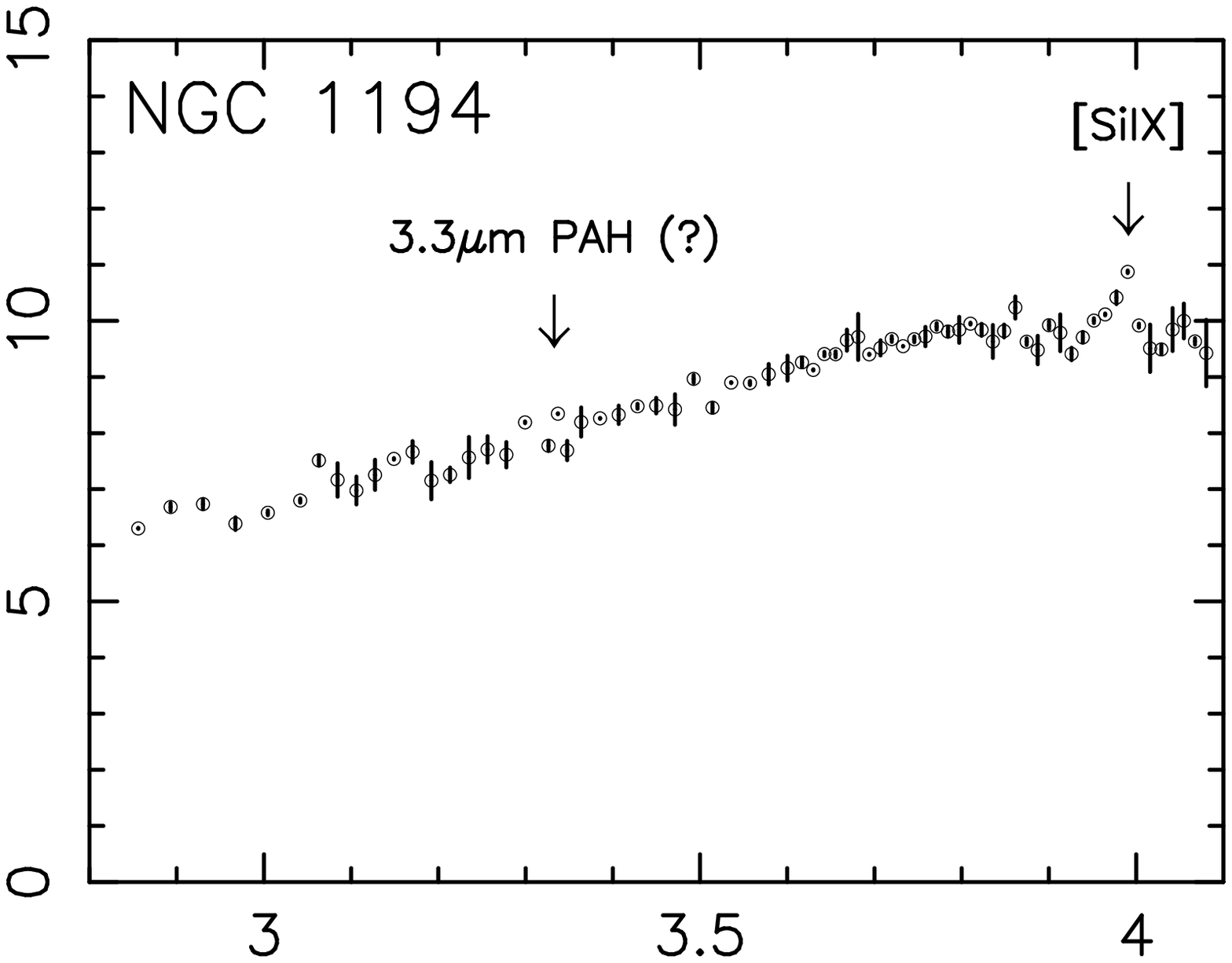}{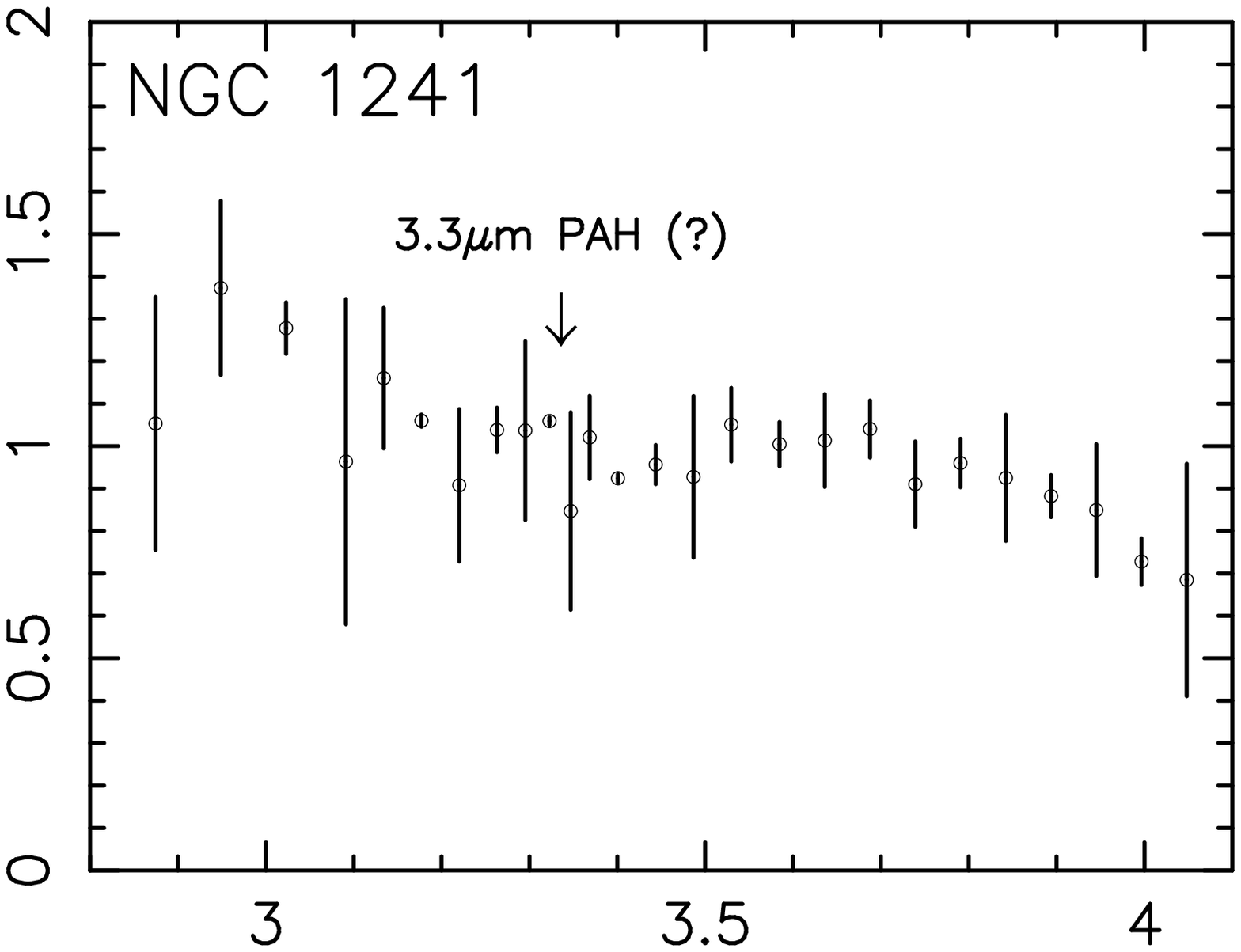}
\end{figure}
\begin{figure} 
\plottwo{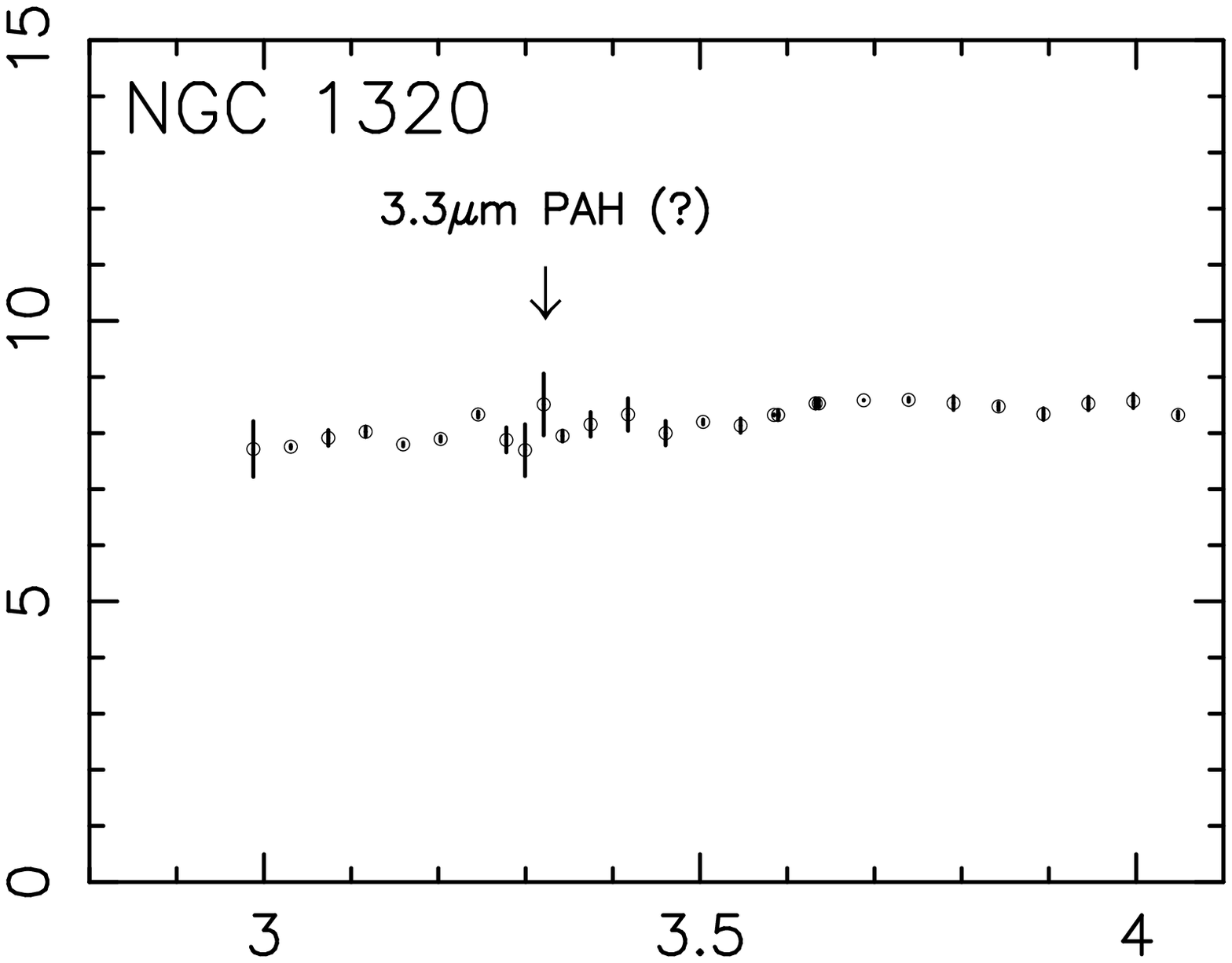}{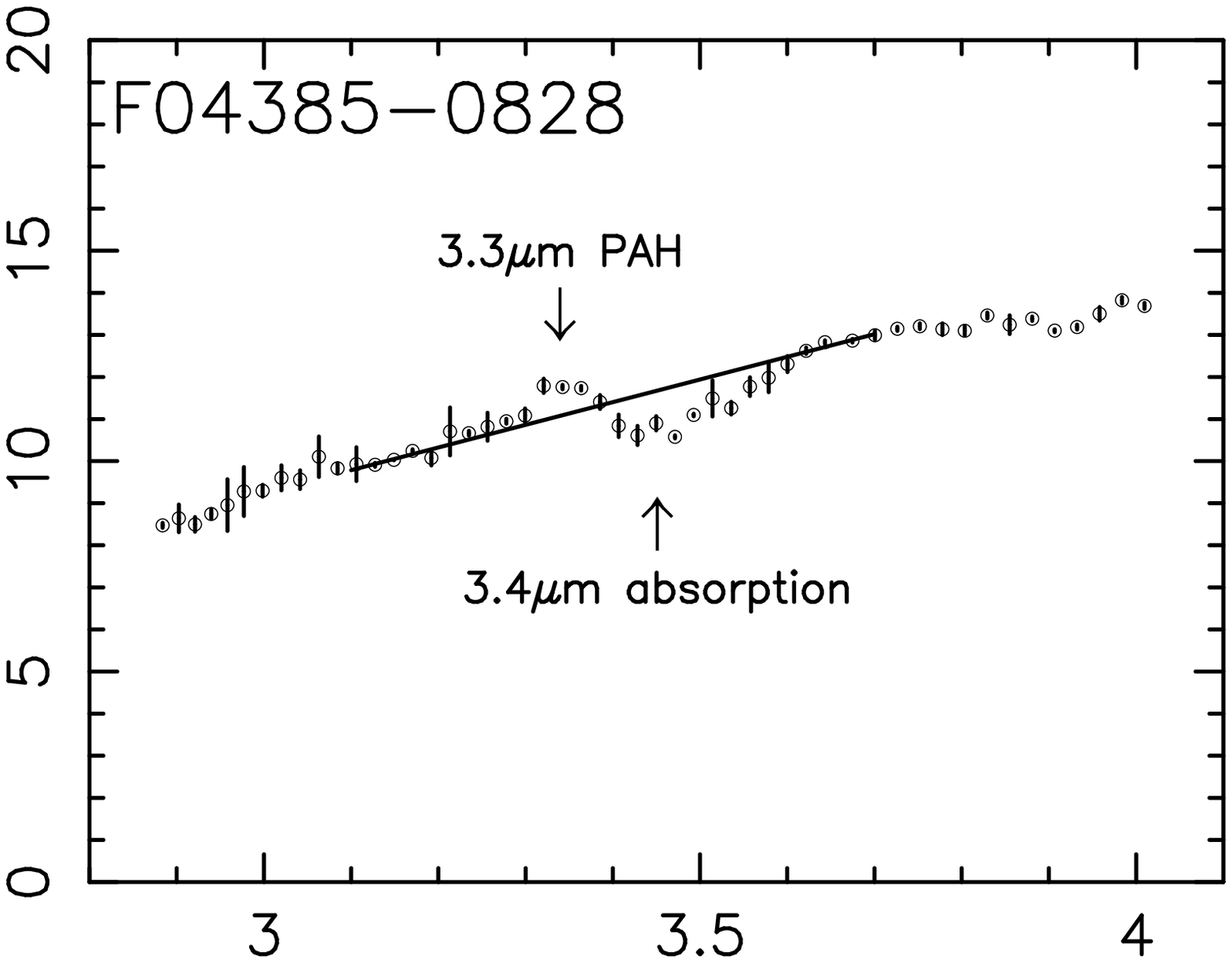}
\end{figure}
\begin{figure} 
\plottwo{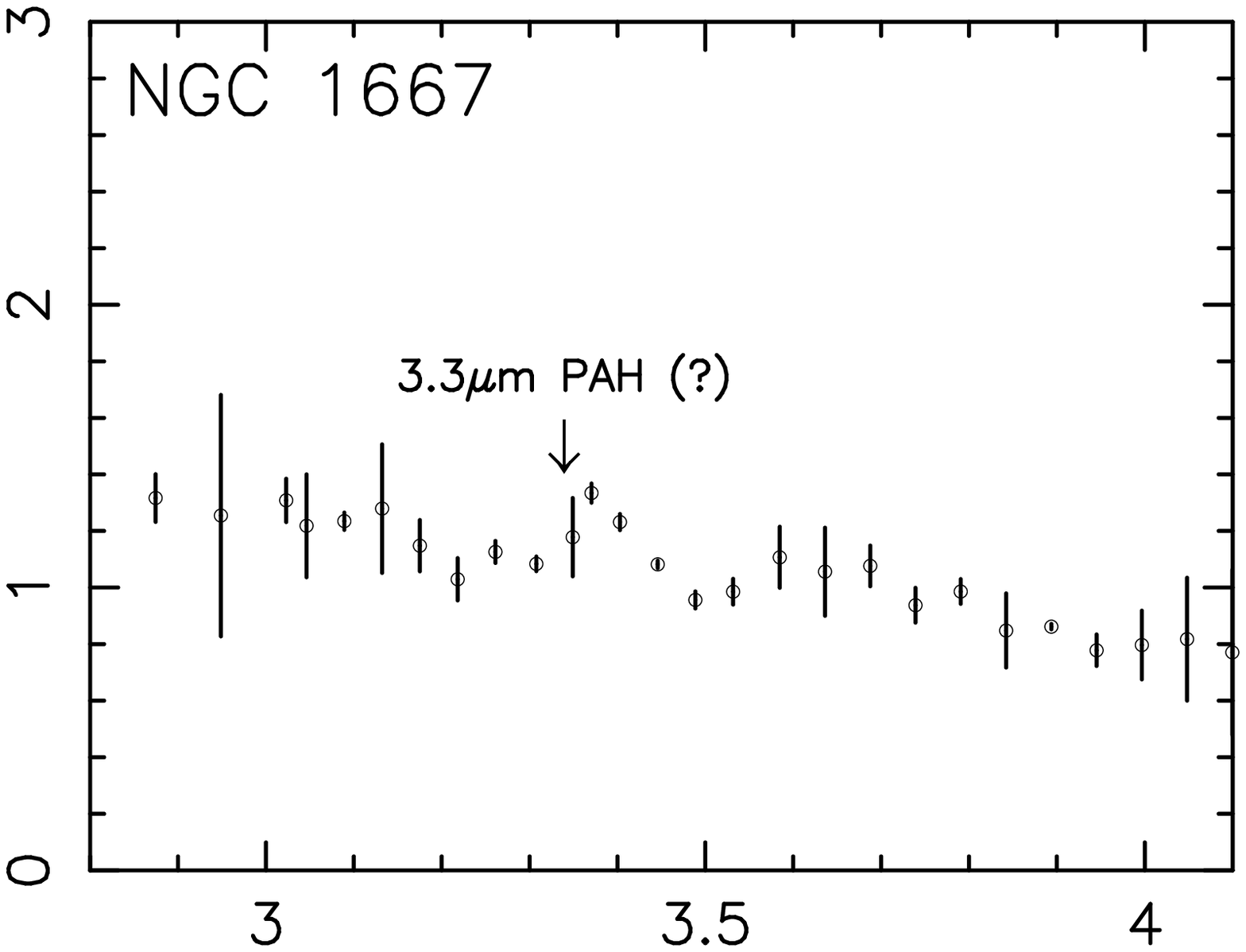}{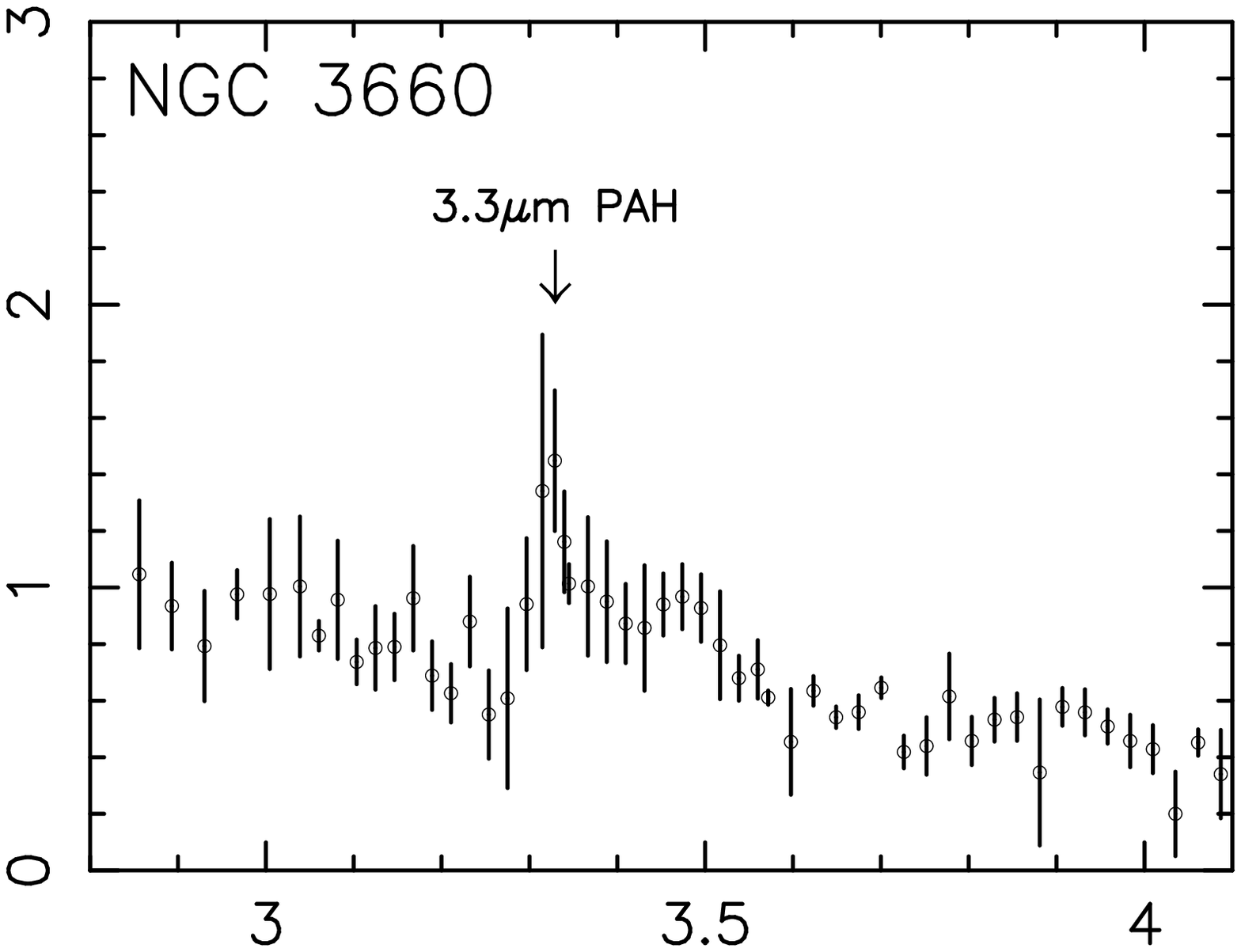}
\end{figure}

\clearpage

\begin{figure} 
\plottwo{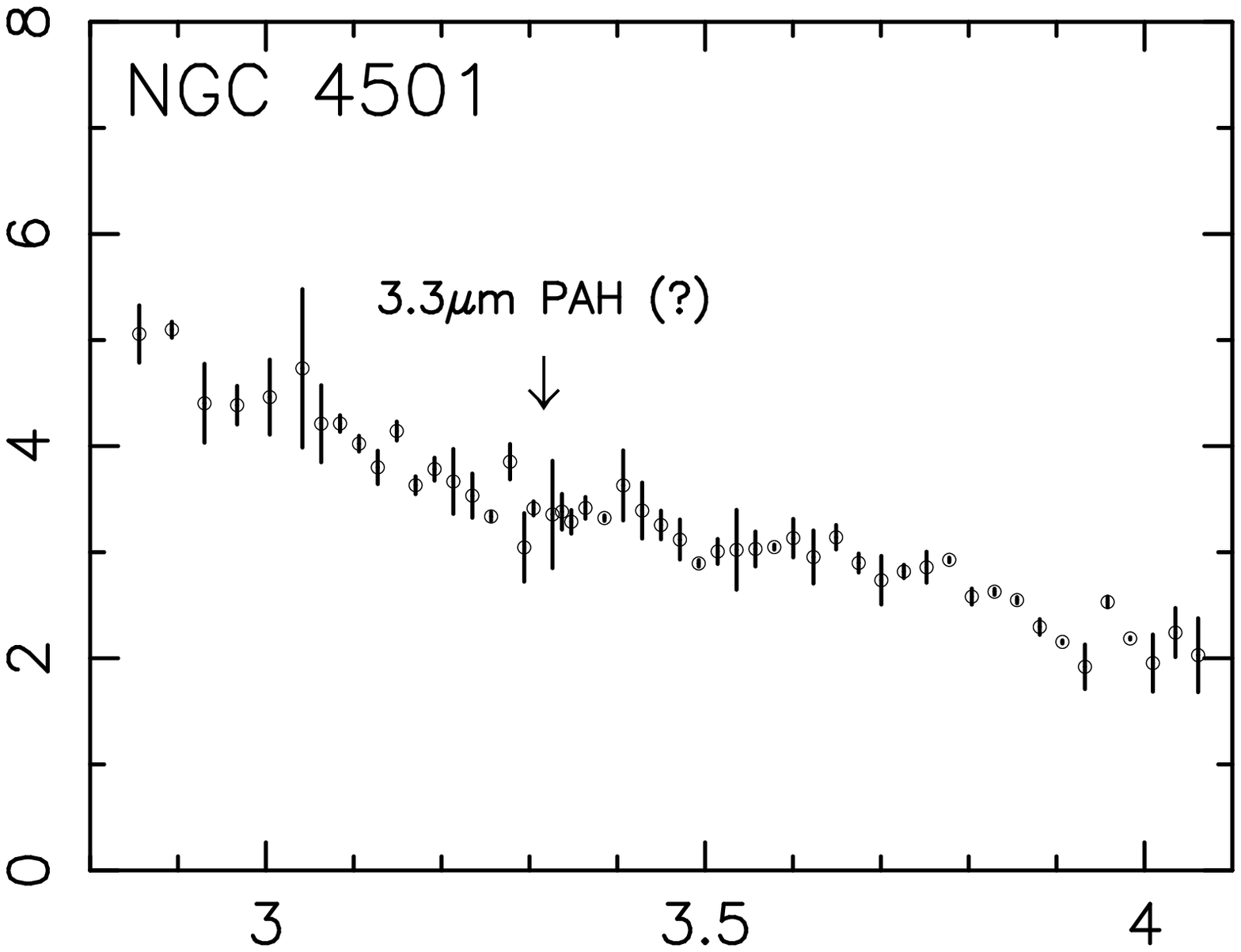}{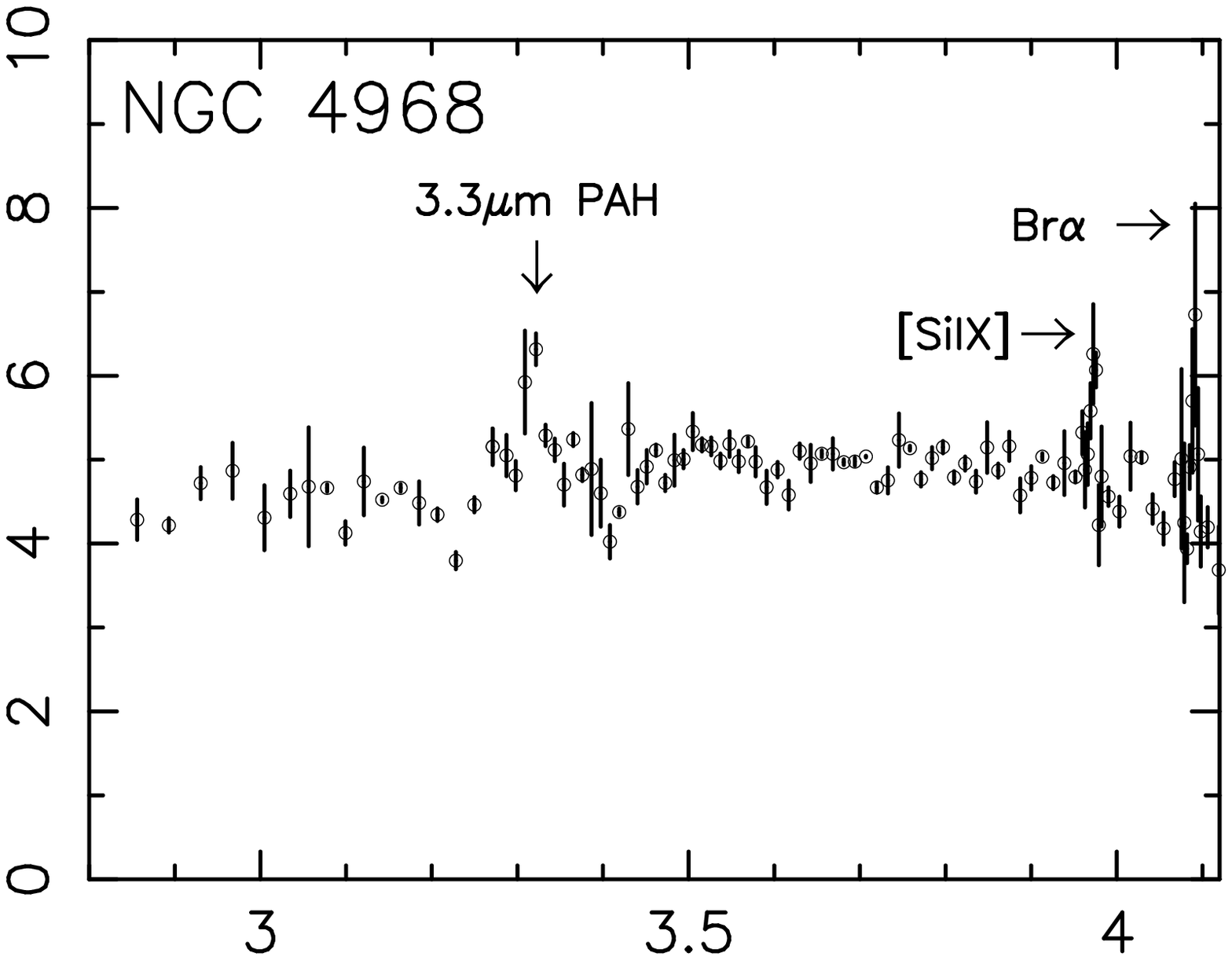}
\end{figure}
\begin{figure} 
\plottwo{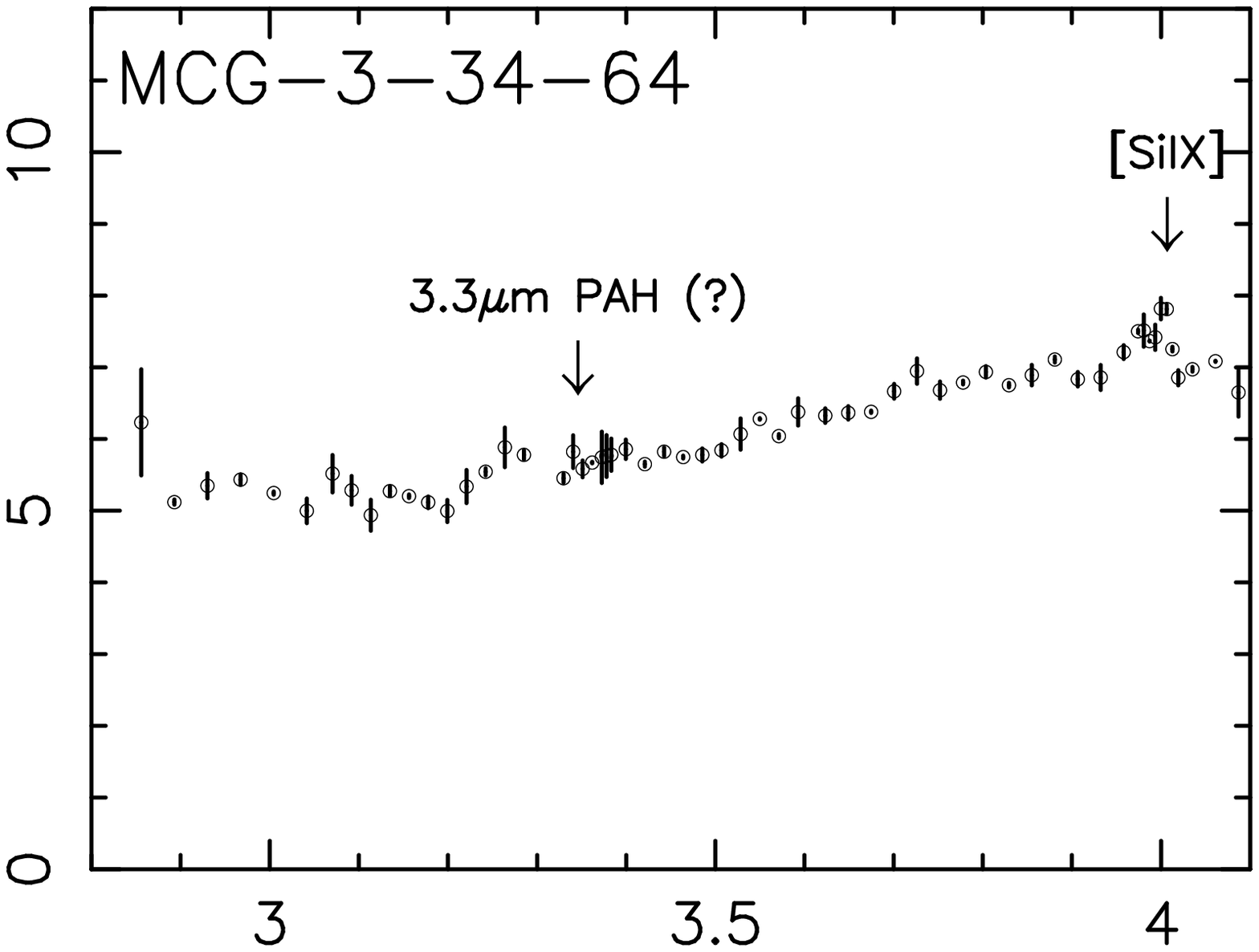}{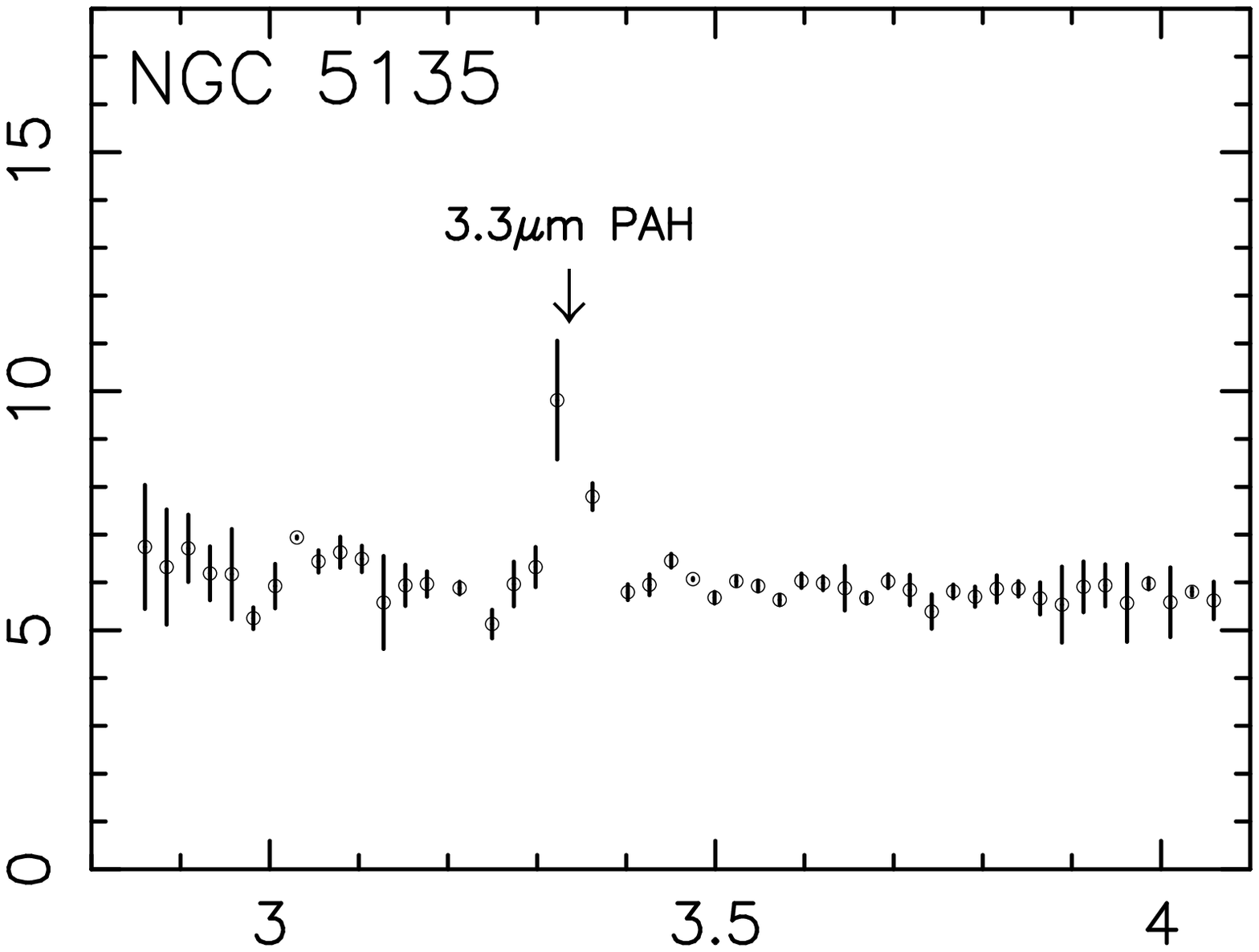}
\end{figure}
\begin{figure} 
\plottwo{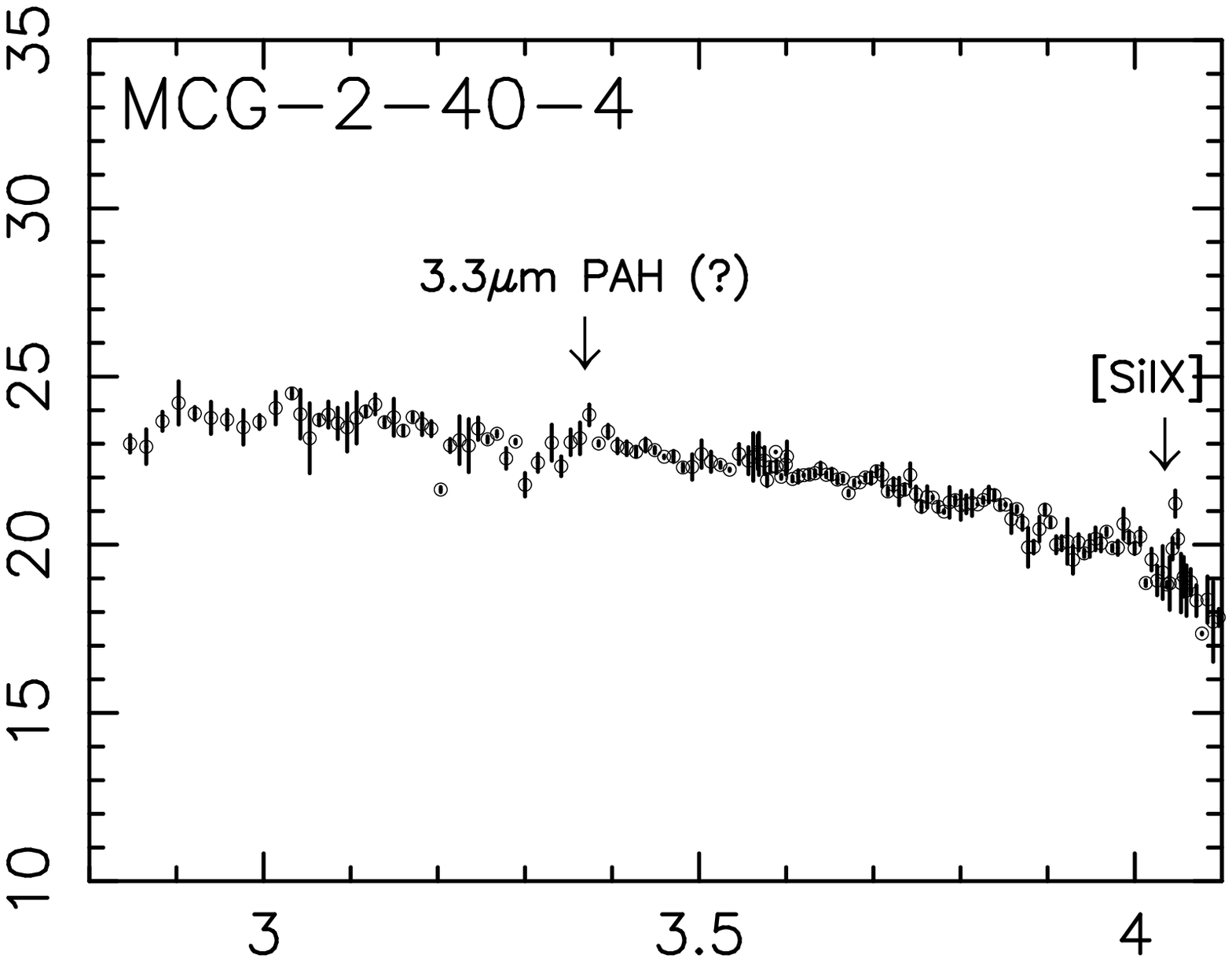}{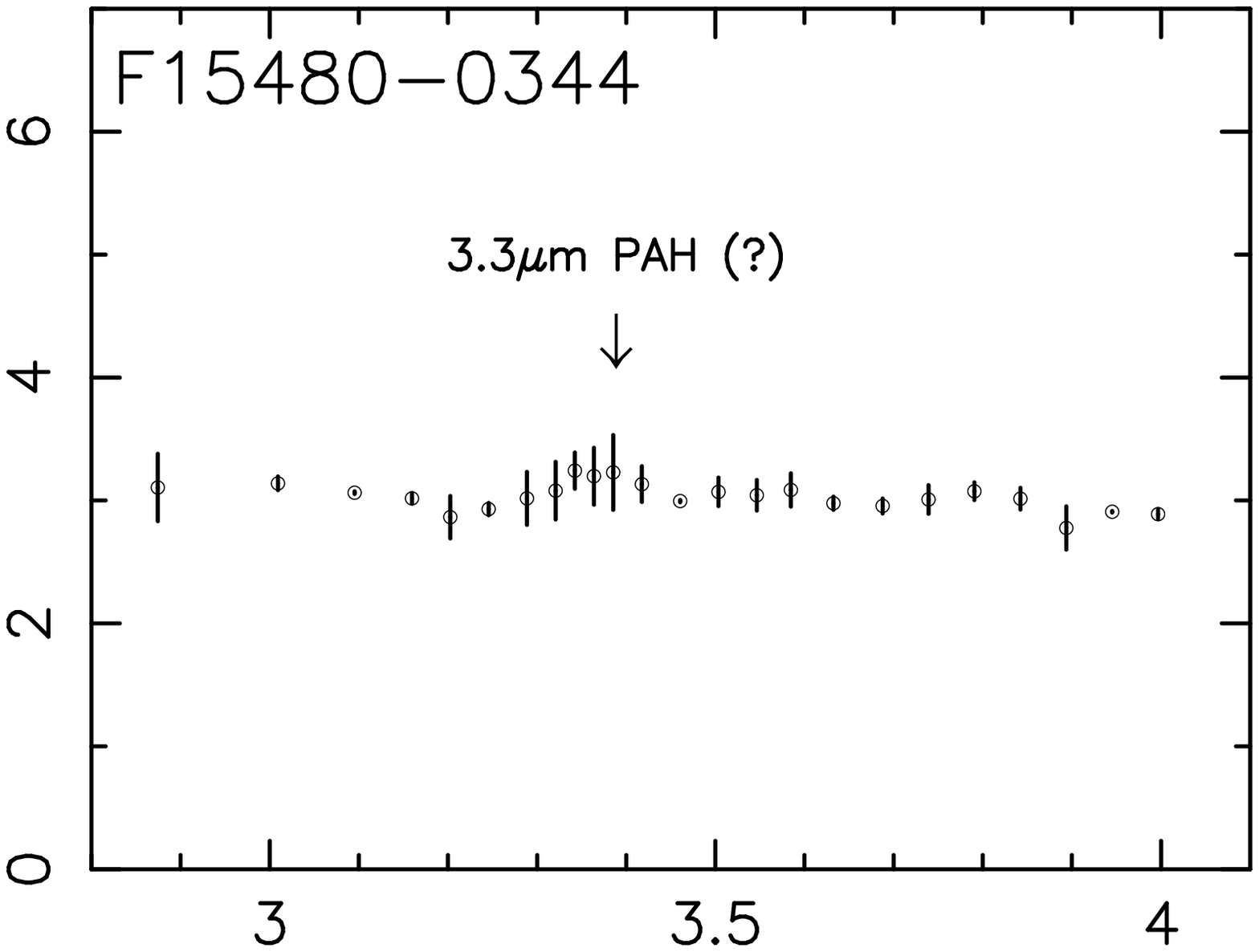}
\end{figure}

\clearpage

\begin{figure}
%\end{figure}
%\begin{figure} 
%\epsscale{0.5}
\plottwo{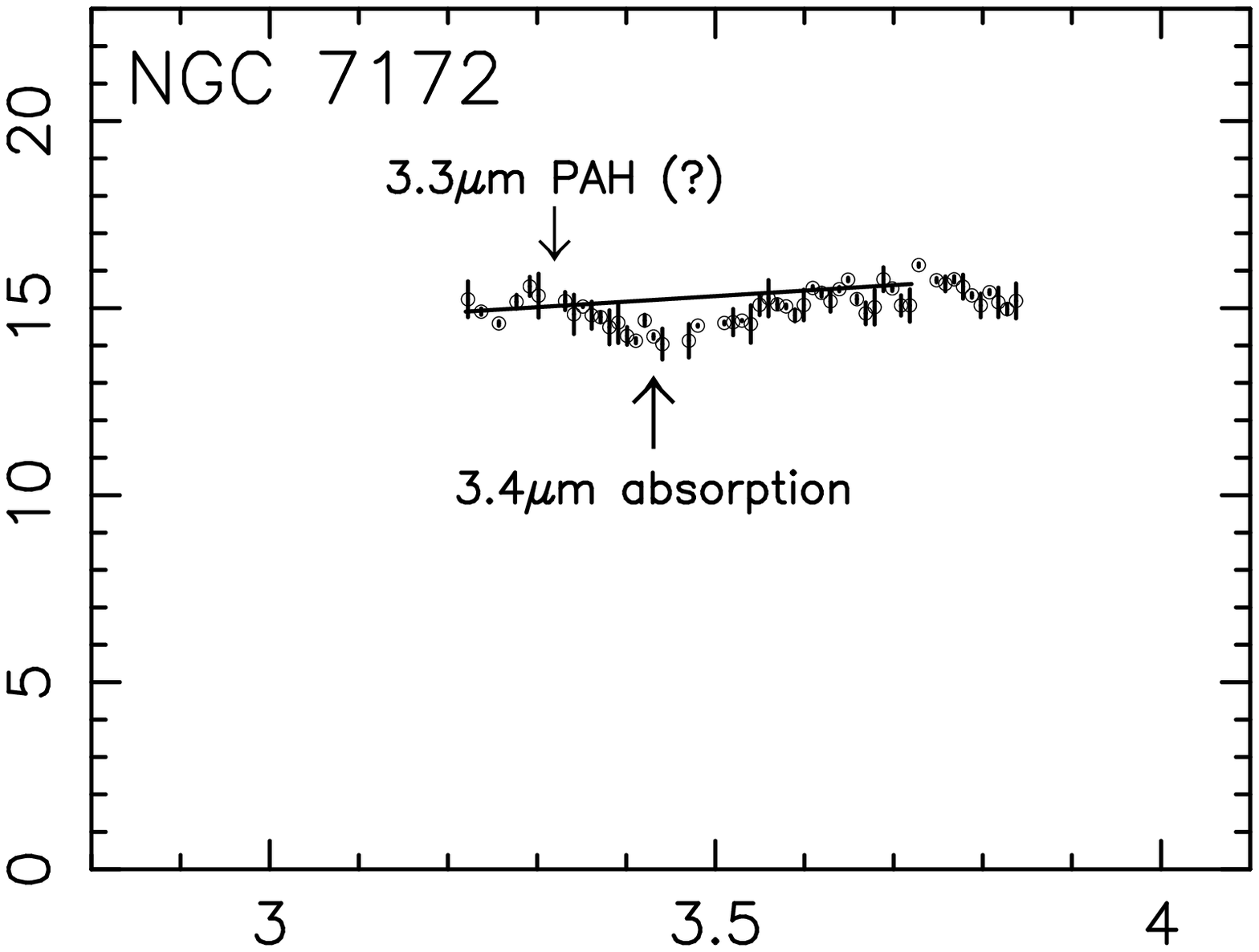}{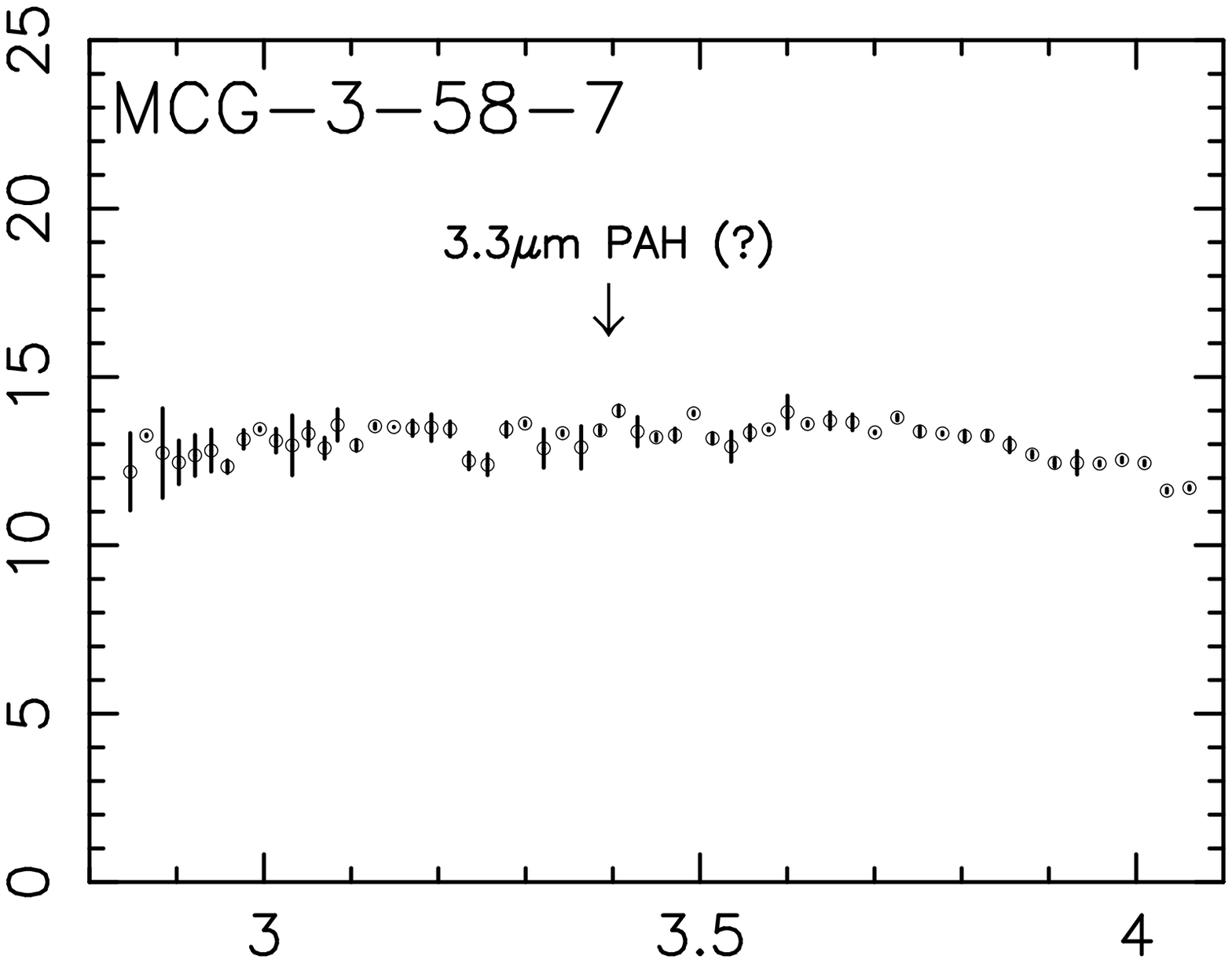}
\end{figure}

\begin{figure}
\caption{Infrared 2.8--4.1 $\mu$m spectra of the 32 Seyfert 2 nuclei.
The abscissa and ordinate are the observed wavelength in $\mu$m and
F$_{\lambda}$ in 10$^{-15}$ W m$^{-2}$ $\mu$m$^{-1}$, respectively.
For NGC 4388, F04385$-$0828, and NGC 7172, the solid lines are the
adopted continuum levels, with respect to which the optical depths of
the absorption feature at 3.1 $\mu$m or 3.4 $\mu$m were measured. 
\label{fig1}}
\end{figure}

\clearpage

\begin{figure}
\epsscale{1.0}
\plottwo{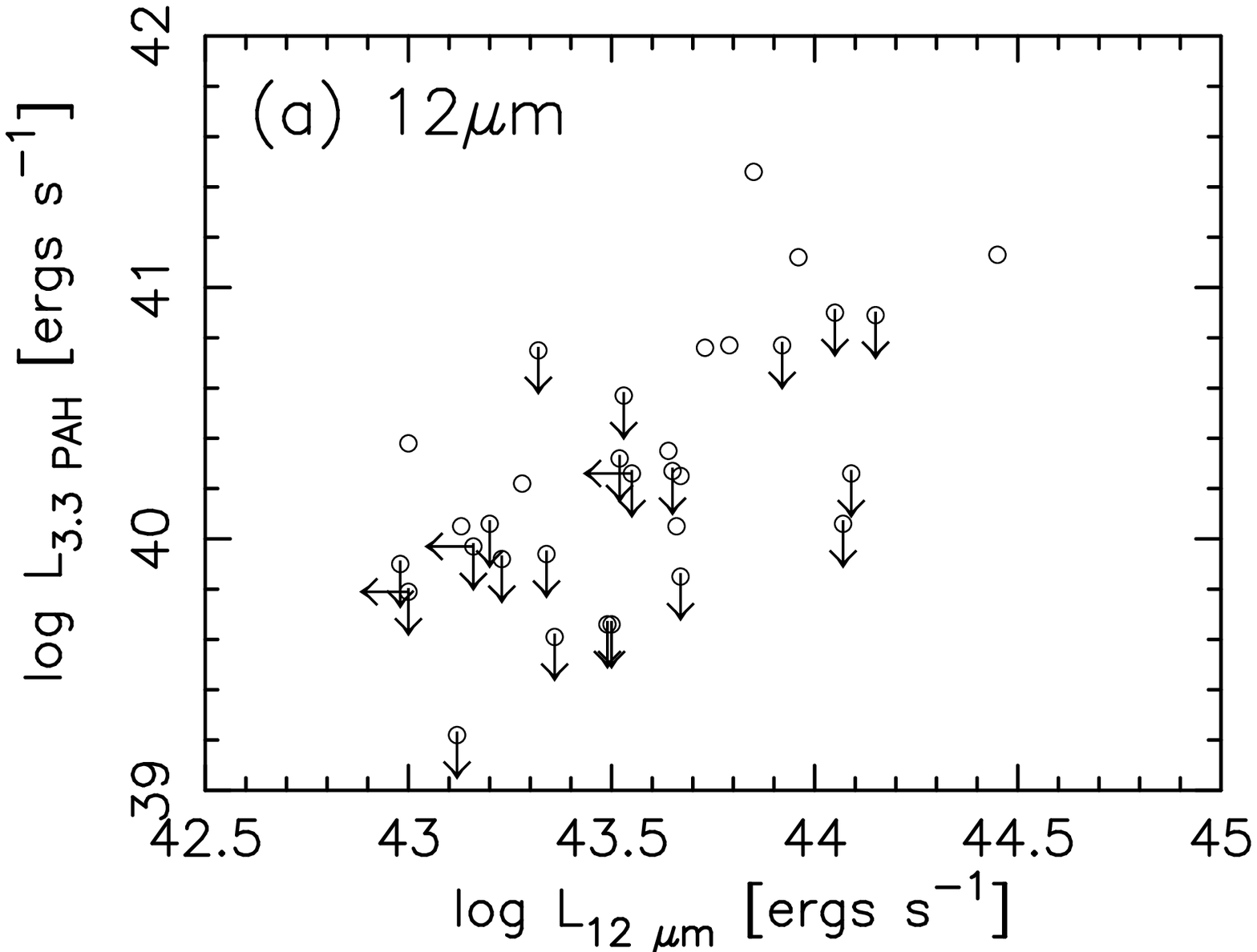}{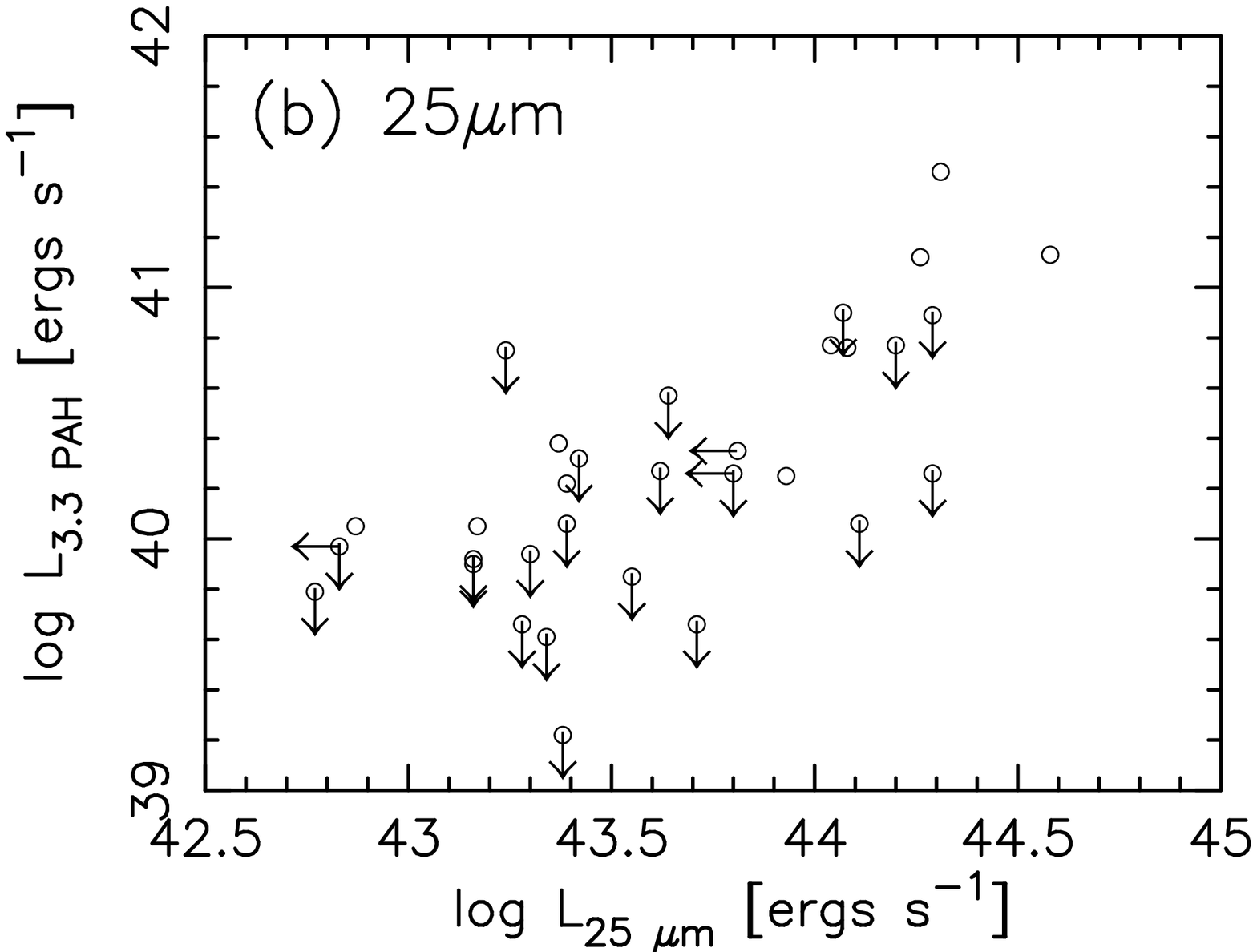}
%\end{figure}
%\begin{figure} 
%\epsscale{0.5}
\plotone{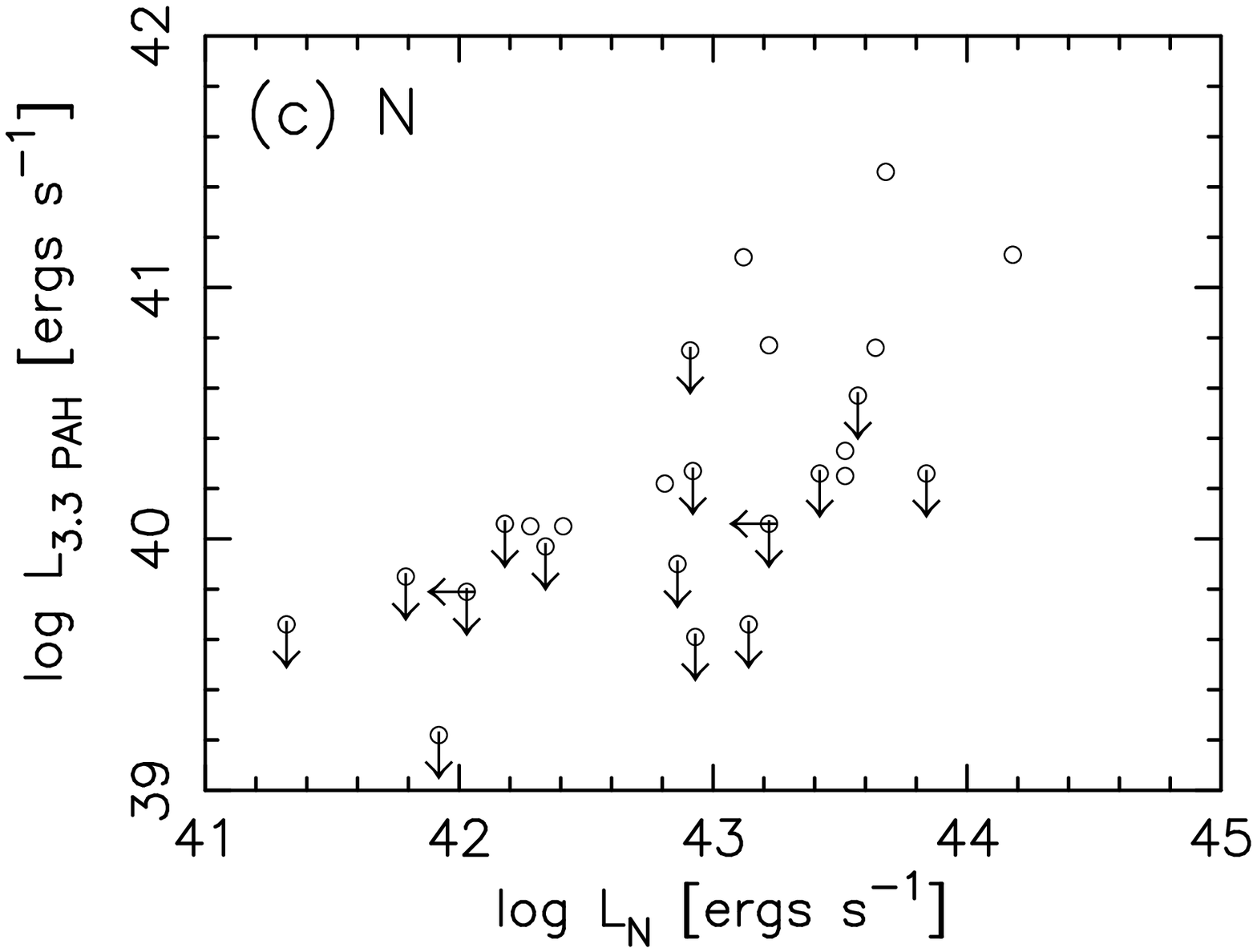}
\end{figure}

\begin{figure}
\caption{
{\it (a)}: The comparison of the {\it IRAS} 12 $\mu$m luminosity,
defined as $\nu$L$_{\nu}$ (abscissa), and 3.3 $\mu$m PAH emission
luminosity detected inside our slit spectra (ordinate). 
{\it (b)}: Same as (a), but the abscissa is {\it IRAS} 25 $\mu$m luminosity. 
{\it (c)}: Same as (a), but the abscissa is $N$-band (10.6 $\mu$m) luminosity
measured with ground-based aperture photometry. 
\label{fig2}}
\end{figure}

\end{document}